%% file: Hybrid_Precoding.tex
\documentclass[journal,twocolumn]{IEEEtran}
\usepackage{amsfonts}
\usepackage{amsmath}
\usepackage{amssymb}
\usepackage{algorithm}
\usepackage{algorithmic}
\usepackage{bm}
\usepackage{cite}
\usepackage{color}
\usepackage{dsfont}
\usepackage{enumerate}
\usepackage{epsfig}
\usepackage{epstopdf}
\usepackage{graphicx}
\usepackage{mathrsfs}
\usepackage{mathtools}
\usepackage{relsize}
\usepackage[T1]{fontenc}
\usepackage{caption}
\usepackage{etoolbox}
\input{mathdef}

\DeclarePairedDelimiterX\set[1]\lbrace\rbrace{#1}
\graphicspath{{./}{Figures/}}

\newtheorem{theorem}{Theorem}%[section]
\newtheorem{lemma}{Lemma}
\newtheorem{proposition}{Proposition}
\newtheorem{corollary}{Corollary}

\begin{document}
\title{Hybrid Precoding For Millimeter Wave MIMO Systems: A Matrix Factorization Approach}

\author{Juening~Jin, Yahong~Rosa~Zheng,~\IEEEmembership{Fellow,~IEEE,} ~Wen~Chen,~\IEEEmembership{Senior Member,~IEEE,} ~Chengshan~Xiao,~\IEEEmembership{Fellow,~IEEE}

\thanks{The work of Y. R. Zheng and C. Xiao was supported in part by US National Science Foundation under Grants ECCS-1231848, ECCS-1408316 and ECCS-1539316. The work of W. Chen was supported in part by National Natural Science Foundation of China under Grant 61671294, STCSM project under Grant 16JC1402900 and 17510740700, Natural Science Foundation of Guangxi Province under Grant 2015GXNSFDA139037, National Science and Technology Major Project under Grant 2017ZX03001002-005 and 2018ZX03001009-002. This work has been carried out while J. Jin was visiting Missouri University of Science and Technology. Part of the material in this paper was presented at the IEEE Globecom, Singapore, 2017.}
\thanks{J. Jin is with the Department of Electronic Engineering, Shanghai Jiao Tong University, Shanghai 200240, China (E-mail:
jueningjin@gmail.com).}
\thanks{Y. R. Zheng is with the Department of Electrical and Computer Engineering,
 Missouri University of Science and Technology, Rolla, MO 65409, USA
 (E-mail: zhengyr@mst.edu).}
\thanks{W. Chen is with the Shanghai Key Laboratory of Navigation and Location
Based Services, Shanghai Jiao Tong University, Shanghai 200240, China (E-mail: wenchen@sjtu.edu.cn).}
\thanks{C. Xiao is with the Department of Electrical and Computer Engineering,
  Lehigh University, Bethlehem, PA 18015, USA
 (E-mail: xiaoc@lehigh.edu).}
}
\maketitle
\begin{abstract}
This paper investigates the hybrid precoding design for millimeter wave (mmWave) multiple-input multiple-output (MIMO) systems with finite-alphabet inputs. The precoding problem is a joint optimization of analog and digital precoders, and we treat it as a matrix factorization problem with power and constant modulus constraints. Our work presents three main contributions: First, we present a sufficient condition and a necessary condition for hybrid precoding schemes to realize unconstrained optimal precoders exactly when the number of data streams $N_\mathrm{s}$ satisfies $N_\mathrm{s}\!=\!\min\{\mathrm{rank}(\bH),N_\mathrm{rf}\}$, where $\bH$ represents the channel matrix and $N_\mathrm{rf}$ is the number of radio frequency (RF) chains. Second, we show that the coupled power constraint in our matrix factorization problem can be removed without loss of optimality. Third, we propose a Broyden-Fletcher-Goldfarb-Shanno (BFGS)-based algorithm to solve our matrix factorization problem using gradient and Hessian information. Several numerical results are provided to show that our proposed algorithm outperforms existing hybrid precoding algorithms.
\end{abstract}

\section{Introduction}
Millimeter wave (mmWave) multiple-input multiple-output (MIMO) communication is a promising technology for future generation cellular systems to address the wireless spectrum crunch. It makes use of the mmWave band from 30 GHz to 300 GHz, which implies a much wider bandwidth than current cellular systems operating in microwave bands. Moreover, a short wavelength of radio signals in the mmWave band enables large number of antennas to be equipped in transceivers, and this allows for applying massive multiple-input multiple-output (MIMO) technique in mmWave communication systems.

For conventional MIMO systems, linear precoding is utilized to maximize the data rate, and it is implemented in the digital domain by the unconstrained optimal precoder. However, the implementation of unconstrained optimal precoders requires one radio frequency (RF) chain per antenna, which will result in prohibitive cost and power consumption in mmWave MIMO systems. To address this issue, a hybrid precoding scheme has been proposed for mmWave MIMO systems to reduce the number of RF chains \cite{el2014spatially,zhang2014achieving,yu2016alternating,sohrabi2016hybrid,rusu2016low,rajashekar2016hybrid,rajashekar2017iterative}. This scheme divides the linear precoder into analog and digital precoders, which are implemented in analog and digital domains, respectively. The digital precoder is realized by a small amount of RF chains, and the analog precoder is realized by phase shifters. Due to the property of phase shifters, each entry of the analog precoder satisfies the constant modulus constraint. These nonconvex constant modulus constraints form a major barrier for hybrid precoding design.

Several hybrid precoding algorithms have been proposed for mmWave MIMO systems \cite{el2014spatially,zhang2014achieving,yu2016alternating,sohrabi2016hybrid,rusu2016low,rajashekar2016hybrid,rajashekar2017iterative}. The work in \cite{el2014spatially} first formulated the hybrid precoding problem as a matrix factorization problem, and then applied the orthogonal
matching pursuit (OMP) algorithm to find near-optimal analog and digital precoders. In \cite{yu2016alternating}, the authors utilized the formulation proposed in \cite{el2014spatially}, and then employed a manifold based alternating minimization algorithm to design hybrid precoders. References \cite{rusu2016low} and \cite{rajashekar2017iterative} introduced and analyzed low complexity hybrid precoding algorithms based on the matrix factorization. There were also some studies on how to achieve the performance of unconstrained optimal precoders with hybrid precoding schemes \cite{zhang2014achieving,sohrabi2016hybrid}, yet requiring the number of RF chains to be twice as much as the number of data streams.

Most existing works on hybrid precoding assume Gaussian inputs, which are rarely realized in practice. It is well known that practical systems utilize finite-alphabet inputs, such as phase-shift keying (PSK) or quadrature amplitude modulation (QAM). Furthermore, it has been shown that precoding designs under Gaussian inputs are quite suboptimal for practical systems with finite-alphabet inputs \cite{guo2005mutual,lozano2006optimum,xiao2008mutual,perez2010mimo,xiao2011globally,zeng2012low,jin2017linear}. A unified framework for linear precoding design under finite-alphabet inputs has been proposed in \cite{jin2017generalized}. Recently, the authors in \cite{rajashekar2016hybrid} presented an iterative gradient ascent algorithm for mmWave MIMO systems with finite-alphabet inputs. In each iteration, the gradient ascent algorithm updated the unconstrained precoder using gradient information, and then it employed a heuristic way to partition the unconstrained precoder into analog and digital precoders. Simulation results illustrated that the gradient ascent algorithm can achieve up to 0.4 bps/Hz gains compared to the Gaussian inputs scenario.

\subsection{Contributions}
In this paper, we investigate the hybrid precoding design for mmWave MIMO systems with finite-alphabet inputs. The contributions of this paper are summarized as follows:
\begin{itemize}
  \item We first provide a sufficient condition under which hybrid precoding schemes can realize any unconstrained optimal precoders exactly. When the sufficient condition does not hold, we also present a necessary condition for hybrid precoding to achieve the performance of unconstrained optimal precoders.
  \item We prove that the power constraint in the hybrid precoding problem \eqref{P2} can be removed without loss of local and/or global optimality. This result greatly simplifies the precoding design, and it enable us to design an efficient algorithm for the hybrid precoding problem.
  \item We present closed form expressions for gradient and Hessian of the hybrid precoding problem. Then we utilize these information to design a BFGS-based algorithm. The proposed algorithm outperforms existing hybrid precoding algorithms.
\end{itemize}

\subsection{Notations}
The following notations are adopted throughout the paper: Boldface lowercase letters, boldface uppercase letters, and calligraphic letters are used to denote vectors, matrices and sets, respectively. The real and complex number fields are denoted by $\mathds{R}$ and $\mathds{C}$, respectively. The superscripts $(\cdot)^{\mathrm{T}}$, $(\cdot)^{*}$ and $(\cdot)^\mathrm{H}$ stand for transpose, conjugate, and conjugate transpose operations, respectively. $\trace(\cdot)$ is the trace of a matrix; $\|\!\cdot\!\|$ denotes the Euclidean norm of a vector; $\|\cdot\|_F$ represents the Frobenius norm of a matrix; $E_{\bx}(\cdot)$ represents the statistical expectation with respect to $\bx$; $\bX_{kl}$ represents the $(k,l)$-th element of $\bX$; $\bI$ and $\bm{0}$ denote an identity matrix and a zero matrix, respectively, with appropriate dimensions; $\bX\!\succeq\!\bm{0}$ denotes a positive semidefinite matrix; $\otimes$ and $\circ$ are Kronecker and Hadamard matrix products, respectively; $\mathcal{I}(\cdot)$ represents the mutual information; $\Re$ and $\Im$ are the real and image parts of a complex value; $\log(\cdot)$ is used for the base two logarithm.

\section{System Model and Problem Formulation}
In this section, we present system and channel models for mmWave MIMO systems, and then formulate the hybrid precoding design as a matrix factorization problem. Finally, we briefly introduce a few notations on complex matrix derivatives.
\subsection{System Model}
Consider a point-to-point mmWave MIMO system, where a transmitter with $N_\mathrm{t}$ antennas sends $N_\mathrm{s}$ data streams to a receiver with $N_\mathrm{r}$ antennas. The number of RF chains at the transmitter is $N_\mathrm{rf}$, which satisfies $N_\mathrm{s}\!\leq\! N_\mathrm{rf}\!\leq\! N_\mathrm{t}$. We consider the hybrid precoding scheme, where $N_\mathrm{s}$ data streams are first precoded using a digital precoder, and then shaped by an analog precoder. The received baseband signal $\by\!\in\! \mathds{C}^{N_\mathrm{r}\times 1}$ can be written as
\begin{align}
\by=\bH\bF_{\!\textrm{RF}}\bF_{\!\textrm{BB}}\bx\!+\!\bn
\end{align}
where $\bH\!\in\! \mathds{C}^{N_\mathrm{r}\times N_\mathrm{t}}$ is the channel matrix; $\bF_{\!\textrm{RF}}\!\in\! \mathds{F}^{N_\mathrm{t}\times N_\mathrm{rf}}$ is the analog precoder with $\mathds{F}\!=\!\big\{f\big\|f|\!=\!\frac{1}{\sqrt{N_\mathrm{t}}}\big\}$ being the constant modulus set; $\bF_{\!\textrm{BB}}\!\in\! \mathds{C}^{N_\mathrm{rf}\times N_\mathrm{s}}$ is the digital precoder; $\bx\!\in\! \mathds{C}^{N_\mathrm{s}\times 1}$ is the input data vector and $\bn\!\in\!\mathds{C}^{N_\mathrm{r}\times 1}$ is the independent and identically distributed (i.i.d.) circularly symmetric complex Gaussian noise with zero-mean and covariance $\sigma^2\bI$.

Suppose that the channel $\bH$ is known at both the transmitter and receiver, and each entry of the input data vector $\bx$ is uniformly distributed from a given constellation set with cardinality $M$. Then the input-output mutual information is given by \cite{xiao2011globally}
\begin{align}\label{MI}
\mathcal{I}(\bx;\by)\!=\!N_\mathrm{s}\log M
\!-\!\frac{1}{M^{N_\mathrm{s}}}\sum_{m=1}^{M^{N_\mathrm{s}}} \!E_{\bn}\Bigg\{\!\log\!\sum_{k=1}^{M^{N_\mathrm{s}}}e^{-d_{mk}}\Bigg\}
\end{align}
where $d_{mk}\!=\!\sigma^{-2}(\|\bH\bF_{\!\textrm{RF}}\bF_{\!\textrm{BB}}(\bx_m\!-\!\bx_k)\!+\!\bn\|^2\!-\!\|\bn\|^2)$, with $\bx_m$ and $\bx_k$ being two possible input data vectors from $\bx$.

\subsection{Channel Model}
The mmWave MIMO channel can be characterized by standard multipath models. Suppose the number of physical propagation paths between the transmitter
and the receiver is $L$. Each path $\ell$ is described by three parameters: complex gain $\alpha_\ell$, angle of arrival $\theta_{\mathrm{r},\ell}$ and angle of departure $\theta_{\mathrm{t},\ell}$. The angles $\{\theta_{\mathrm{r},\ell}\}_{\ell=1}^L$ and $\{\theta_{\mathrm{t},\ell}\}_{\ell=1}^L$ are i.i.d. uniformly distributed over $[0,2\pi)$, and the complex gains $\{\alpha_\ell\}_{\ell=1}^L$ are i.i.d. complex Gaussian distributed with zero-mean and unit-variance. Under this model, the channel matrix $\bH$ is given by \cite[ch. 7.3.2]{tse2005fundamentals}
\begin{align}\label{MCM}
\bH\!=\!\sqrt{\frac{N_\mathrm{r}N_\mathrm{t}}{L}}\sum_{\ell=1}^{L}\alpha_\ell\ba(\theta_{\mathrm{r},l})\ba(\theta_{\mathrm{t},l})^H
\end{align}
where $\ba(\theta_{\mathrm{t},l})$ and $\ba(\theta_{\mathrm{r},l})$ are array steering vectors of the transmit and receive antenna arrays. In this paper, the transmitter and receiver adopt uniform linear arrays, whose array steering vector $\ba(\theta)$ is given by
\begin{align}
\ba(\theta)\!=\!\frac{1}{\sqrt{N}}\Big[1, e^{-\jmath\frac{2\pi}{\lambda}d\sin\theta},...,e^{-\jmath\frac{2\pi}{\lambda}d(N-1)\sin\theta}\Big]^T
\end{align}
where $N$ is the number of antenna element, $\lambda$ is the wavelength of the carrier frequency and $d\!=\!\frac{1}{2}\lambda$ is the antenna spacing.

The channel in \eqref{MCM} can be rewritten in a more compact form as
\begin{align}\label{MCMC}
\bH\!=\!\bA_\mathrm{r}\diag(\bm{\alpha})\bA_\mathrm{t}^{\!H}
\end{align}
where $\bm{\alpha}\!=\![\alpha_1,...,\alpha_L]^T$, $\bA_\mathrm{r}\!\in\! \mathds{C}^{N_\mathrm{r}\times L}$ and $\bA_\mathrm{t}\!\in\! \mathds{C}^{N_\mathrm{t}\times L}$ are array steering matrices with constant modulus entries, given by
\begin{align}
&\bA_\mathrm{r}\!=\!\big[\ba(\theta_{\mathrm{r},1}),...,\ba(\theta_{\mathrm{r},L})\big]\\
&\bA_\mathrm{t}\!=\!\big[\ba(\theta_{\mathrm{t},1}),...,\ba(\theta_{\mathrm{t},L})\big]. \label{TASM}
\end{align}
Note that $\bA_\mathrm{t}$ is a full rank matrix when the angles $\{\theta_{\mathrm{t},l}\}_{l=1}^L$ are distinct \cite{el2014spatially}, and this event occurs with probability one because $\{\theta_{\mathrm{t},l}\}$ are drawn independently from the uniform distribution. Similarly, $\bA_\mathrm{r}$ and $\diag(\bm{\alpha})$  are also full rank matrices with probability one. Therefore, the rank of $\bH$ is given by
\begin{align}\label{rank}
\mathrm{rank}(\bH)\!=\!\min\{L,N_\mathrm{r},N_\mathrm{t}\}.
\end{align}

\subsection{Problem Formulation}
A fundamental approach for hybrid precoding design is to maximize the input-output mutual information under the power and constant modulus constraints. Suppose that the mmWave receiver can optimally decode data using the received signal $\by$, then the hybrid precoding problem is formulated as
\begin{equation}\label{P1}
\begin{aligned}
& \underset{\bF_{\!\textrm{RF}}\in\mathcal{U},\bF_{\!\textrm{BB}}}{\mathrm{maximize}}
& & \quad\quad \mathcal{I}(\bx;\by)\\
& \mathrm{subject \;to}
& & \trace\big(\bF_{\!\textrm{BB}}^H\bF_{\!\textrm{RF}}^H\bF_{\!\textrm{RF}}\bF_{\!\textrm{BB}}\big)\!\leq\! P
\end{aligned}
\end{equation}
where $\mathcal{I}(\bx;\by)$ is given in \eqref{MI}, $P$ is the transmit power constraint and $\mathcal{U}\!=\!\mathds{F}^{N_\mathrm{t}\times N_\mathrm{rf}}$ is the feasible set of analog precoders. It is challenging to solve problem \eqref{P1} directly due to two reasons: First, problem \eqref{P1} is nonconvex because both $\mathcal{I}(\bx;\by)$ and $\mathcal{U}$ are neither convex nor concave with respect to $(\bF_{\!\textrm{RF}},\bF_{\!\textrm{BB}})$. Second, iterative algorithms for problem \eqref{P1} have to evaluate the objective function $\mathcal{I}(\bx;\by)$ in each iteration, which can be very costly because $\mathcal{I}(\bx;\by)$ has no closed form expressions.

To mitigate these difficulties and simplify the precoding design, we adopt the following matrix factorization formulation \cite{el2014spatially}, where hybrid precoders $(\bF_{\!\textrm{RF}},\bF_{\!\textrm{BB}})$ are found by approximating the unconstrained optimal precoder $\bF_{\!\mathrm{opt}}$, i.e.,
\begin{equation}\label{P2}
\begin{aligned}
& \underset{\bF_{\!\textrm{RF}}\in \mathcal{U},\bF_{\!\textrm{BB}}}{\mathrm{minimize}}
& & \|\bF_{\!\mathrm{opt}}\!-\!\bF_{\!\textrm{RF}}\bF_{\!\textrm{BB}}\|_{F}^2\\
& \mathrm{subject \;to}
& & \trace\big(\bF_{\!\textrm{BB}}^H\bF_{\!\textrm{RF}}^H\bF_{\!\textrm{RF}}\bF_{\!\textrm{BB}}\big)\!\leq\! P.
\end{aligned}
\end{equation}
The unconstrained optimal precoder $\bF_{\!\mathrm{opt}}$ is given by \cite{xiao2011globally,zeng2012low}
\begin{equation}
\begin{aligned}
\bF_{\!\mathrm{opt}}\!=& \underset{\bF\in \mathcal{F}}{\mathrm{maximize}}
& & \mathcal{I}(\bx;\by)
\end{aligned}
\end{equation}
where $\mathcal{F}\!=\!\big\{\bF\big|\trace(\bF^H\bF)\!\leq\! P\big\}$.

\subsection{Preliminaries on Complex Matrix Derivatives}
The problems investigated in this paper are nonlinear optimization with complex matrix variables, thus we briefly introduce a few definitions on complex matrix derivatives. For a univariate function $f(x)\!:\! \mathds{C}\rightarrow \mathds{R}$, the definition of the complex derivative is given in \cite{hjorungnes2007complex}:
\begin{align}
\frac{\partial f}{\partial x}\!\triangleq &\frac{1}{2} \bigg[\frac{\partial f}{\partial \Re(x)}-j\frac{\partial f}{\partial \Im(x)}\bigg]\\
\frac{\partial f}{\partial x^*}\!\triangleq &\frac{1}{2} \bigg[\frac{\partial f}{\partial \Re(x)}+j\frac{\partial f}{\partial \Im(x)}\bigg].
\end{align}
For a multivariate function $f(\bX)\!:\! \mathds{C}^{n \times r}\rightarrow \mathds{R}$, the partial derivatives with respect to $\bX$ and $\bX^*$ are matrices
\begin{align}
\frac{\partial f}{\partial \bX}\!\triangleq\! \bigg[\frac{\partial f}{\partial \bX_{k\ell}}\bigg] \quad \mathrm{and}\quad \frac{\partial f}{\partial \bX^*}\!\triangleq\! \bigg[\frac{\partial f}{\partial \bX^*_{k\ell}}\bigg]
\end{align}
where $\bX_{k\ell}$ denotes the $(k,\ell)$-th element of $\bX$. In addition, the complex gradient matrix $\nabla_{\!\scriptscriptstyle\bX}f(\bX)$ is defined as
\begin{align}
\nabla_{\!\scriptscriptstyle\bX}f(\bX)\!\triangleq\! \frac{\partial f}{\partial \bX^*}.
\end{align}
Let $\bX_1\!\in\!\{\bX,\bX^*\}$ and $\bX_2\!\in\!\{\bX,\bX^*\}$, then the complex Hessian of $f(\bX)$ with respect to $\bX_1$ and $\bX_2$ is defined in \cite{hjorungnes2007complex}:
\begin{align}
\mathcal{H}_{\bX_1,\bX_2} f\!\triangleq\! \frac{\partial}{\partial \mathrm{vec}^T(\bX_1)}\bigg[\frac{\partial f}{\partial \mathrm{vec}^T(\bX_2)}\bigg]^T.
\end{align}

\section{Structures of the Hybrid Precoding Problem}
In this section, we first present a sufficient condition and a necessary condition, under which hybrid precoding schemes can realize any unconstrained optimal precoder exactly. Then we prove that the power constraint $\trace\big(\bF_{\!\textrm{BB}}^H\bF_{\!\textrm{RF}}^H\bF_{\!\textrm{RF}}\bF_{\!\textrm{BB}}\big)\!\leq\! P$ in problem \eqref{P2} can be removed without loss of local and/or global optimality.
\subsection{Optimality of Hybrid Precoding Schemes}
The hybrid precoding scheme offers a tradeoff between performance gain and hardware complexity, and its performance is bounded by the unconstrained optimal precoder. When the hybrid precoding scheme can realize any unconstrained optimal precoder exactly, it is an \emph{optimal} scheme. Then a fundamental question arises:
\begin{itemize}
\item Question 1: under what conditions can hybrid precoding schemes realize unconstrained optimal precoders exactly?
\end{itemize}
In other words, we want to find necessary and/or sufficient conditions, under which there exist $(\bF_{\!\textrm{RF}},\bF_{\!\textrm{BB}})$ such that $\bF_{\!\textrm{RF}}\!\in\! \mathcal{U}$ and $\bF_{\!\mathrm{opt}}\!=\!\bF_{\!\textrm{RF}}\bF_{\!\textrm{BB}}$. The best known result related to this question was shown in \cite{zhang2014achieving} and \cite{sohrabi2016hybrid}. It states that when the number of data streams $N_\mathrm{s}$ satisfies $N_\mathrm{s}\!\leq\!\frac{1}{2}N_{\mathrm{rf}}$, we can construct analog and digital precoders to realize any unconstrained optimal precoder with dimensions $N_\mathrm{t}\!\times\! N_\mathrm{s}$. However, this result sacrifices the number of data streams to satisfy $\bF_{\!\mathrm{opt}}\!=\!\bF_{\!\textrm{RF}}\bF_{\!\textrm{BB}}$. In order to achieve the maximum degree of freedom, we should transmit $\mathrm{min}\{\mathrm{rank}(\bH),N_\mathrm{rf}\}$ data streams rather than $\frac{1}{2} N_\mathrm{rf}$ data streams. This motivates us to reconsider Question 1 under $N_\mathrm{s}\!=\!\mathrm{min}\{\mathrm{rank}(\bH),N_\mathrm{rf}\}$.

First, we transform Question 1 into another existence problem through the following proposition.
\begin{proposition}
Suppose $\bF_{\!\textrm{RF}}$ is a full rank matrix, then the following two statements are equivalent:
\begin{enumerate}
  \item There exists $(\bF_{\!\textrm{RF}},\bF_{\!\textrm{BB}})$ such that $\bF_{\!\textrm{RF}}\!\in\! \mathcal{U}$ and $\bF_{\!\mathrm{opt}}\!=\!\bF_{\!\textrm{RF}}\bF_{\!\textrm{BB}}$.
  \item There exists a full rank square matrix $\bS\!\in\!\mathds{C}^{N_\mathrm{rf}\times N_\mathrm{rf}}$ such that $\bU_{\textrm{F}}\bS\!\in\! \mathcal{U}$.
\end{enumerate}
Here $\bU_{\textrm{F}}\!\in\!\mathds{C}^{N_\mathrm{t}\times N_\mathrm{rf}}$ is a semi-unitary matrix whose columns are left singular vectors of $\bF_{\!\mathrm{opt}}$.
\end{proposition}
\begin{IEEEproof}
See Appendix A.
\end{IEEEproof}

Based on Proposition 1, our original problem is equivalent to the existence problem of a full rank square matrix $\bS$ satisfying $\bU_{\textrm{F}}\bS\!\in\! \mathcal{U}$. By exploiting the inherent structure of the mmWave MIMO channel, we provide a sufficient condition to guarantee the existence of such full rank matrix $\bS$. The main idea is similar to Theorem 1 of \cite{raghavan2016beamforming}.
%depends on $\bU_\textrm{F}$, which has a close connection with the channel matrix $\bH$. Let the singular value decomposition (SVD) of $\bH$ be
%\begin{align}
%\bH\!=\!\bU_{\textrm{H}}\mathbf{\Sigma}_{\textrm{H}}\bV_{\!\textrm{H}}^H
%\end{align}
%where $\bU_{\textrm{H}}\!\in\!\mathds{C}^{N_{\mathrm{r}}\times N_{\mathrm{t}}}$ is a unitary matrix with left singular vectors, $\mathbf{\Sigma}_{\textrm{H}}\!\in\!\mathds{C}^{N_{\mathrm{t}}\times N_{\mathrm{t}}}$ is a diagonal matrix with singular values arranged in decreasing order, and $\bV_{\!\textrm{H}}\!\in\!\mathds{C}^{N_{\mathrm{t}}\times N_{\mathrm{t}}}$ is a unitary matrix with right singular vectors. According to \cite[Proposition 2]{xiao2011globally}, $\bU_{\textrm{F}}$ can always be chosen as the first $N_\mathrm{rf}$ columns of $\bV_{\!\mathrm{H}}$, i.e.,
%\begin{align}
%\bU_\textrm{F}\!=\!\big[{\bV_{\!\mathrm{H}}}\big]_{\bullet,1:N_\mathrm{RF}}.
%\end{align}
\begin{proposition}
When the number of paths $L$ satisfies $L\!\leq\! \min\{N_\mathrm{r},N_\mathrm{t},N_\mathrm{rf}\}$, there exists a full rank matrix $\bS$ satisfying $\bA_\mathrm{t}\!=\!\bU_\textrm{F}\bS\!\in\!\mathcal{U}$, where $\bA_\mathrm{t}$ is the array steering matrix given in \eqref{TASM}.
\end{proposition}
\begin{IEEEproof}
See Appendix A.
\end{IEEEproof}

Combining Propositions 1 and 2, we conclude that when $L\!\leq\! \min\{N_\mathrm{r},N_\mathrm{t},N_\mathrm{rf}\}$, hybrid precoding schemes can realize any unconstrained optimal precoder $\bF_{\!\mathrm{opt}}$ exactly. However, the sufficient condition in Proposition 2 does not always hold in practice because the number of paths may be greater than the number of RF chains. In the rest of this subsection, we propose a necessary condition for the existence of $\bS$ satisfying $\bU_\mathrm{F}\bS\!\in\!\mathcal{U}$, and the proposed necessary condition is independent of $L$, $N_\mathrm{rf}$, $N_\mathrm{r}$ and $N_\mathrm{t}$.

We first rewrite $\bU_\mathrm{F}\bS\!\in\!\mathcal{U}$ as
\begin{align}\label{S3}
\big[\bU_\mathrm{F}\bs_\ell\bs_\ell^H\bU_\mathrm{F}^H\big]_{kk}\!=\!\frac{1}{N_{\mathrm{t}}},\; k=1,...,N_{\mathrm{t}},\ell=1,...,N_\mathrm{rf}
\end{align}
where $\bs_\ell$ is the $\ell$th column of $\bS$. Combining condition \eqref{S3} and $\mathrm{rank}(\bS)\!=\!N_\mathrm{rf}$, the original problem is equivalent to the existence of $N_\mathrm{rf}$ linear independent solutions $\{\bs_\ell\}_{l=1}^{N_\mathrm{rf}}$ to the following system of quadratic equations:
\begin{align}\label{S4}
\big[\bU_\mathrm{F}\bs\bs^H\bU_\mathrm{F}^H\big]_{kk}\!=\!\frac{1}{N_{\mathrm{t}}},\; k=1,...,N_{\mathrm{t}}.
\end{align}
Unfortunately, problem \eqref{S4} is intractable because checking the existence of solutions to a general quadratic system is NP-hard \cite{chen2015solving}. Instead, we investigate necessary conditions for the existence of solutions to \eqref{S4}.

The main idea is to transform \eqref{S4} into a linear system by semidefinite programming. Define $\bZ\!=\!N_\mathrm{t}\bs\bs^H$, the quadratic system \eqref{S4} can be written as
\begin{align}\label{S5}
\big[\bU_\mathrm{F}\bZ\bU_\mathrm{F}^H\big]_{kk}\!=\!1,\; \forall k,\;\bZ\!\succeq\!\bm{0},\;\mathrm{rank}(\bZ)\!=\!1.
\end{align}
Furthermore, according to
\begin{align}
\mathrm{vec}\big(\bU_\mathrm{F}\bZ\bU_\mathrm{F}^H\big)\!=\!\big(\bU_\mathrm{F}^*\otimes \bU_\mathrm{F}\big)\mathrm{vec}(\bZ)
\end{align}
equations \eqref{S5} is expressed more compactly as
\begin{align}\label{S6}
\bK_\mathrm{F}\mathrm{vec}(\bZ)\!=\!\mathbf{1},\;\bZ\!\succeq\!\bm{0},\;\mathrm{rank}(\bZ)\!=\!1
\end{align}
where the $k$th row of $\bK_\mathrm{F}$ is chosen as the $\big[(k-1)N_\mathrm{t}+k\big]$th row of $\bU_\mathrm{F}^*\otimes \bU_\mathrm{F}$. Through some standard algebraic manipulations, we can express $\bK_\mathrm{F}$ as
\begin{align}
\bK_\mathrm{F}\!=\!\Big[\diag(\bu_1^*)\bU_\mathrm{F},...,\diag\big(\bu_{N_\mathrm{rf}}^*\big)\bU_\mathrm{F}\Big]
\end{align}
where $\bu_\ell$ represents the $\ell$th column of $\bU_\mathrm{F}$.

The main barrier for solving equations \eqref{S6} is the nonlinear constraints $\bZ\!\succeq\! \bm{0}$ and $\mathrm{rank}(\bZ)\!=\!1$, which restrict solutions of $\bK_\mathrm{F}\mathrm{vec}(\bZ)\!=\!\mathbf{1}$ with a certain structure. Therefore, we first relax the nonlinear constraints and focus on the linear system $\bK_\mathrm{F}\mathrm{vec}(\bZ)\!=\!\mathbf{1}$. Clearly, if equations \eqref{S6} has $N_\mathrm{rf}$ linear independent solutions, then $\bK_\mathrm{F}\mathrm{vec}(\bZ)\!=\!\mathbf{1}$ should have at least $N_\mathrm{rf}$ linear independent solutions. Based on this observation, the following proposition provides a necessary condition for the existence of a full rank $\bS$ such that $\bU_\textrm{F}\bS\!\in\!\mathcal{U}$.
\begin{proposition}
If there exist a full rank square matrix $\bS$ satisfying $\bU_\textrm{F}\bS\!\in\!\mathcal{U}$, then
\begin{align}
\mathrm{rank}(\bK_\mathrm{F})\!\leq\! N_\mathrm{rf}^2\!-\!N_\mathrm{rf}\!+\!1
\end{align}
\end{proposition}
\begin{IEEEproof}
See Appendix A.
\end{IEEEproof}

Note that we can compute $\mathrm{rank}(\bK_\mathrm{F})$ without the knowledge of $\bF_{\!\mathrm{opt}}$ because its left singular vectors $\bU_\textrm{F}$ can always be chosen as the first $N_\mathrm{rf}$ columns of $\bV_{\textrm{H}}$, with $\bV_{\textrm{H}}\!\in\!\mathds{C}^{N_\mathrm{t}\times N_\mathrm{t}}$ being the right singular vectors of $\bH$ \cite[Proposition 2]{xiao2011globally}. Therefore, when the transmitter has perfect channel state information, it can construct $\bK_\textrm{F}$ and check whether $\mathrm{rank}(\bK_\mathrm{F})\!\leq\! N_\mathrm{rf}^2\!-\!N_\mathrm{rf}\!+\!1$ holds. If the necessary condition does not hold, then hybrid precoding schemes cannot realize unconstrained optimal precoders exactly.

When the sufficient condition in Proposition 2 does not hold, $\bK_\mathrm{F}$ is usually a full rank matrix. In this case, we derive the minimum number of RF chains required for hybrid precoding to achieve the performance of unconstrained optimal precoders.
\begin{corollary}
When $\bK_\mathrm{F}$ is a full rank matrix, it requires at least $\sqrt{N_\mathrm{t}-\frac{3}{4}}+\frac{1}{2}$ RF chains for hybrid precoding schemes to realize unconstrained optimal precoders exactly.
\end{corollary}
\begin{IEEEproof}
Since $\bK_\mathrm{F}$ is a full rank matrix, $\mathrm{rank}(\bK_\mathrm{F})\!=\!\min\{N_\mathrm{t},N_\mathrm{rf}^2\}$. Inserting $\mathrm{rank}(\bK_\mathrm{F})$ into $\mathrm{rank}(\bK_\mathrm{F})\!\leq\! N_\mathrm{rf}^2-N_\mathrm{rf}+1$ and using quadratic formula, we obtain
\begin{align}
N_\mathrm{rf}\!\geq\!\sqrt{N_\mathrm{t}-\frac{3}{4}}+\frac{1}{2}.
\end{align}
This completes the proof.
\end{IEEEproof}

\subsection{Structures of the Matrix Factorization Formulation}
Given the unconstrained optimal precoder $\bF_{\!\mathrm{opt}}$, the matrix factorization problem \eqref{P2} belongs to the class of polynomial optimization: The objective function $\|\bF_{\!\mathrm{opt}}\!-\!\bF_{\!\textrm{RF}}\bF_{\!\textrm{BB}}\|_{F}^2$ is a convex quartic function with respect to matrix variables $(\bF_{\!\textrm{RF}},\bF_{\!\textrm{BB}})$, the power constraint $\trace\big(\bF_{\!\textrm{BB}}^H\bF_{\!\textrm{RF}}^H\bF_{\!\textrm{RF}}\bF_{\!\textrm{BB}}\big)\!\leq\! P$ is a convex quartic constraint, and the constant modulus constraints $\mathcal{U}$ are nonconvex quadratic equality constraints. Such a problem is nonconvex due to the nonconvexity of $\mathcal{U}$, and theoretical challenges of problem \eqref{P2} are listed as follows:
\begin{enumerate}
\item The optimization variables $\bF_{\!\textrm{RF}}$ and $\bF_{\!\textrm{BB}}$ are coupled through the power constraint. Therefore, we cannot deploy the alternating minimization approach which requires separate variables in constraints. If we jointly optimize $(\bF_{\!\textrm{RF}},\bF_{\!\textrm{BB}})$, the difficulty also lies in handing the coupled feasible region of problem \eqref{P2}.
\item More importantly, the bilinear mapping $(\bF_{\!\textrm{RF}},\bF_{\!\textrm{BB}})\!\mapsto\! \bF_{\!\textrm{RF}}\bF_{\!\textrm{BB}}$ is not a one-to-one mapping, thus $(\bF_{\!\textrm{RF}},\bF_{\!\textrm{BB}})$ and $(\bF_{\!\textrm{RF}}\mathbf{\Sigma},\mathbf{\Sigma}^{-1}\bF_{\!\textrm{BB}})$ result in the same objective value, where $\mathbf{\Sigma}$ is a diagonal matrix with unit modulus diagonal entries to ensure $\bF_{\!\textrm{RF}}\mathbf{\Sigma}\!\in\!\mathcal{U}$. In other words, we should expect problem \eqref{P2} to have infinite number of local minima and saddle points.
\end{enumerate}

The first issue is fully addressed by the following theorem, which shows the equivalence between problems \eqref{P2} and the following relaxed problem:
\begin{equation}\label{P3}
\begin{aligned}
& \underset{\bF_{\!\textrm{RF}}\in \mathcal{U},\bF_{\!\textrm{BB}}}{\mathrm{minimize}}
& & \|\bF_{\!\mathrm{opt}}\!-\!\bF_{\!\textrm{RF}}\bF_{\!\textrm{BB}}\|_{F}^2.
\end{aligned}
\end{equation}
\begin{theorem}
If $(\hat{\bF}_{\!\textrm{RF}},\hat{\bF}_{\!\textrm{BB}})$ is a KKT point of problem \eqref{P3}, then it satisfies $\trace\big(\hat{\bF}_{\!\textrm{BB}}^H\hat{\bF}_{\!\textrm{RF}}^H\hat{\bF}_{\!\textrm{RF}}\hat{\bF}_{\!\textrm{BB}}\big)\!\leq\! P$.
\end{theorem}
\begin{IEEEproof}
See Appendix A.
\end{IEEEproof}

According to Theorem 1, any KKT point of problem \eqref{P3} satisfies $\trace\big(\bF_{\!\textrm{BB}}^H\bF_{\!\textrm{RF}}^H\bF_{\!\textrm{RF}}\bF_{\!\textrm{BB}}\big)\!\leq\! P$, thus the power constraint can be removed without loss of local and global optimality.

The rest of this paper focuses on solving problem \eqref{P3}. Problem \eqref{P3} is a constant modulus matrix factorization problem where a given matrix $\bF_{\!\mathrm{opt}}$ is factorized into two complex matrices $(\bF_{\!\textrm{RF}}$, $\bF_{\!\textrm{BB}})$ under constant modulus constraints on $\bF_{\!\textrm{RF}}$. Since $(\bF_{\!\textrm{RF}},\bF_{\!\textrm{BB}})\!\mapsto\! \bF_{\!\textrm{RF}}\bF_{\!\textrm{BB}}$ is not a one-to-one mapping, problem \eqref{P3} has infinite number of saddle points, and this issue will be addressed in Section IV.

\section{Constant Modulus Matrix Factorization}
\subsection{Problem Reformulation}
First, we observe that for any given $\bF_{\!\textrm{RF}}$, problem \eqref{P3} is a least square problem
\begin{equation}\label{31}
\begin{aligned}
& \underset{\bF_{\!\textrm{BB}}}{\mathrm{minimize}}
& & \|\bF_{\!\mathrm{opt}}\!-\!\bF_{\!\textrm{RF}}\bF_{\!\textrm{BB}}\|_{F}^2.
\end{aligned}
\end{equation}
Suppose that $\bF_{\!\textrm{RF}}$ has full column rank, then the optimal solution of problem \eqref{31} is
\begin{align}\label{32}
\bF_{\!\textrm{BB}}=\bF_{\!\textrm{RF}}^{+}\bF_{\!\mathrm{opt}}
\end{align}
where $\bF_{\!\textrm{RF}}^{+}\!=\!(\bF_{\!\textrm{RF}}^H\bF_{\!\textrm{RF}})^{-1}\bF_{\!\textrm{RF}}^H$ is the Moore-Penrose pseudoinverse of $\bF_{\!\textrm{RF}}$. Inserting \eqref{32} into problem \eqref{P3}, $\bF_{\!\textrm{BB}}$ is eliminated and we obtain the modified problem:
\begin{equation}\label{P4}
\begin{aligned}
& \underset{\bF_{\!\textrm{RF}}\in \mathcal{U}}{\mathrm{minimize}}\quad
f(\bF_{\!\textrm{RF}})\!=\!\|\bF_{\!\mathrm{opt}}\!-\!\bF_{\!\textrm{RF}}\bF_{\!\textrm{RF}}^{+}\bF_{\!\mathrm{opt}}\|_{F}^2.
\end{aligned}
\end{equation}
The following theorem guarantees that problems \eqref{P3} and \eqref{P4} are equivalent.
\begin{theorem}
If $\hat{\bF}_{\!\textrm{RF}}$ is a KKT point of problem \eqref{P4} and $\hat{\bF}_{\!\textrm{BB}}\!=\!\hat{\bF}_{\!\textrm{RF}}^{+}\bF_{\!\mathrm{opt}}$, then $(\hat{\bF}_{\!\textrm{RF}}, \hat{\bF}_{\!\textrm{BB}})$ is a KKT point of problem \eqref{P3}. Furthermore, $\hat{\bF}_{\!\textrm{RF}}$ is a globally optimal solution of problem \eqref{P4} if and only if $(\hat{\bF}_{\!\textrm{RF}}, \hat{\bF}_{\!\textrm{BB}})$ is a globally optimal solution of problem \eqref{P3}.
\end{theorem}
\begin{IEEEproof}
See Appendix B.
\end{IEEEproof}

The benefit of this reformulation is that problem \eqref{P4} can be solved more efficiently because its search space is reduced from $(\bF_{\!\textrm{RF}}, \bF_{\!\textrm{BB}})$ to $\bF_{\!\textrm{RF}}$.

Problem \eqref{P4} involves minimizing a polynomial with nonconvex constant modulus constraints, which is difficult to handle. Note that the constant modulus constraints imply that only the phase of $\bF_{\!\textrm{RF}}$ can be changed. Therefore, instead of using $\bF_{\!\textrm{RF}}$ as the optimization variable, it is more convenient to optimize the phase of $\bF_{\!\textrm{RF}}$ directly. Let the phase of $\bF_{\!\textrm{RF}}$ be $\bm{\Phi}_{\textrm{RF}}$, i.e., $\bF_{\!\textrm{RF}}\!=\!\frac{1}{\sqrt{N_\mathrm{t}}}e^{j\bm{\Phi}_{\textrm{RF}}}$. Using $\bm{\Phi}_{\textrm{RF}}$ as the optimization variable and rewriting $\bF_{\!\textrm{RF}}$ as $\bF_{\!\textrm{RF}}(\bm{\Phi}_{\textrm{RF}})$, we can reformulate problem \eqref{P4} as the following unconstrained minimization problem
\begin{equation}\label{P5}
\begin{aligned}
\underset{\bm{\Phi}_{\textrm{RF}}}{\mathrm{minimize}}\;\; \psi(\bm{\Phi}_{\textrm{RF}})\!=\!\|\bF_{\!\mathrm{opt}}\!-\!\bF_{\!\textrm{RF}}(\bm{\Phi}_{\textrm{RF}})\bF_{\!\textrm{RF}}^+(\bm{\Phi}_{\textrm{RF}})\bF_{\!\mathrm{opt}}\|_{F}^2.
\end{aligned}
\end{equation}

Although \eqref{P5} is a unconstrained problem, it is still not recommended to solve this problem directly because the objective function $\psi(\bm{\Phi}_{\textrm{RF}})$ is ill-behaved: First, $\psi(\bm{\Phi}_{\textrm{RF}})\!=\!\psi(\bm{\Phi}_{\textrm{RF}}\!+\!\bS)$ for any rank one real matrix $\bS$. Thus problem \eqref{P5} has infinite number of local minima and saddle points; Second, the Hessian of $\psi(\bm{\Phi}_{\textrm{RF}})$ at any point $\bm{\Phi}_{\textrm{RF}}$ is a singular matrix. To show this, we expand $\psi(\bm{\Phi}_{\textrm{RF}}\!+\!\bS)$ at $\bm{\Phi}_{\textrm{RF}}$ using Taylor's theorem:
\begin{align}
\psi(\bm{\Phi}_{\textrm{RF}}\!+\!\bS)&\!=\!\psi(\bm{\Phi}_{\textrm{RF}})\!+\!\mathrm{vec}\big[\nabla\psi(\bm{\Phi}_{\textrm{RF}})\big]^T\mathrm{vec}(\bS)\nonumber\\
&\!+\!\frac{1}{2}\mathrm{vec}(\bS)^T\big[\nabla^2\psi(\bm{\Phi}_{\textrm{RF}})\big]\mathrm{vec}(\bS)\!+\!o(\|\mathrm{vec}(\bS)\|^2)\nonumber
\end{align}
where $\nabla\psi(\bm{\Phi}_{\textrm{RF}})$ and $\nabla^2\psi(\bm{\Phi}_{\textrm{RF}})$ are the gradient and Hessian of $\psi(\bm{\Phi}_{\textrm{RF}})$ respectively, and $o(\|\mathrm{vec}(\bS)\|^2)$ is the Peano's form of the reminder. For any nonzero rank one real matrix $\bS$, we have $\psi(\bm{\Phi}_{\textrm{RF}}\!+\!\bS)\!=\!\psi(\bm{\Phi}_{\textrm{RF}})$, which implies
\begin{align}
&\mathrm{vec}(\bS)\!\neq\!\mathbf{0}\nonumber\\
&\mathrm{vec}\big[\nabla\psi(\bm{\Phi}_{\textrm{RF}})\big]^T\mathrm{vec}(\bS)\!=\!0\\ &\big[\nabla^2\psi(\bm{\Phi}_{\textrm{RF}})\big]\mathrm{vec}(\bS)\!=\!\mathbf{0}.\nonumber
\end{align}
Therefore, $\nabla^2\psi(\bm{\Phi}_{\textrm{RF}})$ is a singular matrix.

We address these two issues by restricting the first row of $\bm{\Phi}_{\textrm{RF}}$ being a zero vector. Note that $\bm{\Phi}_{\textrm{RF}}$ can be partitioned into two blocks
\begin{align}
\bm{\Phi}_{\textrm{RF}}\!=\!
\begin{bmatrix}
\br\\
\bR
\end{bmatrix}
\end{align}
where $\br\!\in\!\mathds{R}^{1\times N_\mathrm{rf}}$ is the first row of $\bm{\Phi}_{\textrm{RF}}$, and $\bR\!\in\! \mathds{R}^{(N_\mathrm{t}-1)\times N_\mathrm{rf}}$ is the remaining part of $\bm{\Phi}_{\textrm{RF}}$. If $\br$ is not a zero vector, we can always construct a unique matrix
\begin{align}
\mathbf{\bar{\Phi}}_{\textrm{RF}}\!=\!\bm{\Phi}_{\textrm{RF}}\!-\!\mathbf{1}\br\!=\!
\begin{bmatrix}
\bm{0}\\
\bm{\Phi}
\end{bmatrix}
\end{align}
such that the first row of $\mathbf{\bar{\Phi}}_{\textrm{RF}}$ is a zero vector, and $\psi(\mathbf{\bar{\Phi}}_{\textrm{RF}})\!=\!\psi(\bm{\Phi}_{\textrm{RF}})$. Therefore, we can optimize $\psi(\bm{\Phi}_{\textrm{RF}})$ over a special class of $\bm{\Phi}_{\textrm{RF}}$ satisfying
\begin{align}\label{P61}
\bm{\Phi}_{\textrm{RF}}\!=\!
\begin{bmatrix}
\bm{0}\\
\bm{\Phi}
\end{bmatrix}
\end{align}
where $\mathbf{\Phi}\!\in\! \mathds{R}^{(N_\mathrm{t}-1)\times N_\mathrm{rf}}$. Using $\mathbf{\Phi}$ as the optimization variable, problem \eqref{P5} is further reformulated as
\begin{equation}\label{P6}
\begin{aligned}
& \underset{\mathbf{\Phi}}{\mathrm{minimize}}
& & \varphi(\mathbf{\Phi})\!=\!\psi\bigg\{\!
\begin{bmatrix}
\bm{0}\\
\bm{\Phi}
\end{bmatrix}\!\bigg\}.
\end{aligned}
\end{equation}

\subsection{Gradient and Hessian}
In this subsection, we derive the gradient and Hessian of $\varphi(\mathbf{\Phi})$, which are the foundation for developing numerical algorithms to solve problem \eqref{P6}. Since the gradient and Hessian of $\varphi(\mathbf{\Phi})$ depend on those of $f(\bF_{\!\textrm{RF}})$, we first provide the gradient and Hessian of $f(\bF_{\!\textrm{RF}})$ in the following lemma.
\begin{lemma}
The complex gradient and Hessian matrices of $f(\bF_{\!\textrm{RF}})$ are given by
\begin{align}
\nabla_{\!\bF_{\!\textrm{RF}}} f(\bF_{\!\textrm{RF}})&\!\triangleq\!\frac{\partial f(\bF_{\!\textrm{RF}})}{\partial \bF_{\!\textrm{RF}}^*}\!=\!-\bZ_1\bF_{\!\mathrm{opt}}\bZ_2^H\\
\mathcal{CH}_{\bF_{\!\textrm{RF}}} f(\bF_{\!\textrm{RF}})&\!\triangleq\!
\begin{bmatrix}
\mathcal{H}_{\bF_{\!\textrm{RF}},\bF_{\!\textrm{RF}}^*} f(\bF_{\!\textrm{RF}}) & \mathcal{H}_{\bF_{\!\textrm{RF}}^*,\bF_{\!\textrm{RF}}^*} f(\bF_{\!\textrm{RF}}) \\
\mathcal{H}_{\bF_{\!\textrm{RF}},\bF_{\!\textrm{RF}}} f(\bF_{\!\textrm{RF}}) & \mathcal{H}_{\bF_{\!\textrm{RF}}^*,\bF_{\!\textrm{RF}}} f(\bF_{\!\textrm{RF}})
\end{bmatrix}\nonumber\\
&\!=\!
\begin{bmatrix}
\mathcal{H}_{\bF_{\!\textrm{RF}},\bF_{\!\textrm{RF}}^*} f(\bF_{\!\textrm{RF}}) & \mathcal{H}_{\bF_{\!\textrm{RF}}^*,\bF_{\!\textrm{RF}}^*} f(\bF_{\!\textrm{RF}}) \\
\mathcal{H}^*_{\bF_{\!\textrm{RF}}^*,\bF_{\!\textrm{RF}}^*} f(\bF_{\!\textrm{RF}}) & \mathcal{H}^*_{\bF_{\!\textrm{RF}},\bF_{\!\textrm{RF}}^*} f(\bF_{\!\textrm{RF}})
\end{bmatrix}
\end{align}
where $\bZ_1\!=\!\bI\!-\!\bF_{\!\textrm{RF}}\bF^+_{\!\textrm{RF}}$, $\bZ_2\!=\!\bF^+_{\!\textrm{RF}}\bF_{\!\mathrm{opt}}$, and
\begin{align}
\mathcal{H}_{\bF_{\!\textrm{RF}},\bF_{\!\textrm{RF}}^*} f(\bF_{\!\textrm{RF}})\!=& (\bZ_2\bZ_2^H)^T\!\otimes\! \bZ_1\nonumber\\
&\!-\!\big[(\bF_{\!\textrm{RF}}^H\bF_{\!\textrm{RF}})^{-1}\big]^T\!\otimes\!\bZ_1\bF_{\!\mathrm{opt}}\bF_{\!\mathrm{opt}}^H\bZ_1^H\nonumber
\end{align}
\begin{align}
\mathcal{H}_{\bF_{\!\textrm{RF}}^*,\bF_{\!\textrm{RF}}^*} f(\bF_{\!\textrm{RF}})\!=& \big[(\bZ_1\bF_{\!\mathrm{opt}}\bZ_2^H)^T\otimes(\bF_{\!\textrm{RF}}^{+})^H\big]\bK_{N_{\mathrm{t}},N_\mathrm{rf}}\nonumber\\
&\!+\!\bK_{N_{\mathrm{t}},N_\mathrm{rf}}^T\big[(\bZ_1\bF_{\!\mathrm{opt}}\bZ_2^H)^T\otimes(\bF_{\!\textrm{RF}}^{+})^H\big]^T.\nonumber
\end{align}
Here $\bK_{N_{\mathrm{t}},N_\mathrm{rf}}$ is the commutation matrix satisfying $\mathrm{vec}(\mathrm{d}\bF_{\!\textrm{RF}}^T)\!=\!\bK_{N_{\mathrm{t}},N_\mathrm{rf}}\mathrm{vec}(\mathrm{d}\bF_{\!\textrm{RF}})$.
\end{lemma}
\begin{IEEEproof}
See Appendix B.
\end{IEEEproof}

With the help of Lemma 1, we can compute the gradient $\nabla\varphi(\mathbf{\Phi})$ and Hessian $\nabla^2\varphi(\mathbf{\Phi})$. For any given $\mathbf{\Phi}$, we construct the corresponding $\bm{\Phi}_\textrm{RF}$ in \eqref{P61}. Then $\nabla\varphi(\mathbf{\Phi})$ is obtained by deleting the first row of $\nabla \psi(\bm{\Phi}_\textrm{RF})$, and $\nabla^2\varphi(\mathbf{\Phi})$ is obtained by deleting the $(N_{\mathrm{t}}\ell\!+\!1)$th rows and columns of $\nabla^2 \psi(\bm{\Phi}_\textrm{RF})$, with $\ell\!=\!0,1,...,N_\mathrm{rf}\!-\!1$. The gradient and Hessian of $\psi(\bm{\Phi}_\textrm{RF})$ are given in the following theorem.
\begin{theorem}
The gradient and Hessian matrices of $\psi(\bm{\Phi}_\textrm{RF})$ are given by
\begin{align}
\nabla \psi(\bm{\Phi}_\textrm{RF})&\!=\!2\Im\big[\bG\big]\\
\nabla^2 \psi(\bm{\Phi}_\textrm{RF})&\!=\!2\Re\big[\bM\big]\!-\!2\diag\big\{\mathrm{vec}\big(\Re\big[\bG\big]\big)\big\}
\end{align}
where $\bG\!=\!\nabla_{\!\bF_{\!\textrm{RF}}} f(\bF_{\!\textrm{RF}})\!\circ\! \bF_{\!\textrm{RF}}^*$ and
\begin{align}
\bM\!=&[\mathcal{H}_{\bF_{\!\textrm{RF}},\bF_{\!\textrm{RF}}^*} f(\bF_{\!\textrm{RF}})]\!\circ\!\mathrm{vec}(\bF_{\!\textrm{RF}}^*)\mathrm{vec}(\bF_{\!\textrm{RF}})^T\nonumber\\
&\!-\![\mathcal{H}_{\bF_{\!\textrm{RF}}^*,\bF_{\!\textrm{RF}}^*} f(\bF_{\!\textrm{RF}})]\circ\!\mathrm{vec}(\bF_{\!\textrm{RF}}^*)\mathrm{vec}(\bF_{\!\textrm{RF}})^H.
\end{align}
\end{theorem}
\begin{IEEEproof}
See Appendix B.
\end{IEEEproof}

\subsection{BFGS-based Algorithm}
In this subsection, we propose a Broyden-Fletcher-Goldfarb-Shanno (BFGS)-based method to solve problem \eqref{P6}. The BFGS method is a well-known quasi-Newton algorithm for unconstrained optimization problems. It updates the current solution $\mathbf{\Phi}_{\!n}$ to $\mathbf{\Phi}_{\!n+1}$ by the following rule:
\begin{align}\label{BFGS1}
\mathbf{\Phi}_{\!n+1}\!=\! \mathbf{\Phi}_{\!n}\!+\!\rho_n\bS_{n}
\end{align}
where $\bS_{n}$ is the descent direction, and $\rho_n\!>\!0$ is the stepsize. The descent direction $\bS_{n}$ is given by
\begin{align}\label{BFGS2}
\mathrm{vec}(\bS_{n})\!=\!-\bB_{n}\mathrm{vec}[\nabla \varphi(\mathbf{\Phi}_{\!n})].
\end{align}
Here $\bB_{n}$ is a symmetric positive definite matrix which approximates the inverse of $\nabla^2\varphi(\mathbf{\Phi}_{\!n})$. Note that the positive definiteness of $\bB_{n}$ ensures that $\bS_n$ is a descent direction, i.e.,
\begin{align}
\trace\big[\nabla \varphi(\mathbf{\Phi}_{\!n})^T\bS_n\big]\!=\!-\mathrm{vec}[\nabla \varphi(\mathbf{\Phi}_{\!n})]^T\bB_{n}\mathrm{vec}[\nabla \varphi(\mathbf{\Phi}_{\!n})]\!<\!0.\nonumber
\end{align}
The matrix $\bB_{n}$ is usually updated by the inverse BFGS formula
\begin{align}\label{BFGS3}
\bB_{n+1}\!=\!\bigg(\bI\!-\!\frac{\bs_{n}\by_{\!n}^T}{\by_{\!n}^T\bs_n}\bigg)\bB_{n}\bigg(\bI\!-\!\frac{\bs_{n}\by_{\!n}^T}{\by_{\!n}^T\bs_n}\bigg)^T\!+\!\frac{\bs_{n}\bs_{n}^T}{\by_{\!n}^T\bs_n}
\end{align}
where $\bs_n\!=\!\mathrm{vec}[\mathbf{\Phi}_{\!n+1}\!-\!\mathbf{\Phi}_{\!n}]$ and $\by_{\!n}\!=\!\mathrm{vec}[\nabla \varphi(\mathbf{\Phi}_{\!n+1})\!-\!\nabla \varphi(\mathbf{\Phi}_{\!n})]$. Clearly, $\bB_{n+1}$ will inherit the positive definiteness of $\bB_n$ as long as $\by_{\!n}^T\bs_n\!>\!0$. However, the condition $\by_{\!n}^T\bs_n\!>\!0$ does not hold for general nonconvex problems. In order to ensure the positive definiteness of $\bB_{n+1}$, a cautious update rule for $\bB_{n}$ is proposed\cite{li2001global}
\begin{align}\label{BFGS4}
\bB_{n+1}\!=\!\left\{
\begin{aligned}
&\eqref{BFGS3}\quad\;\; \mathrm{if}\;\frac{\by_{\!n}^T\bs_n}{\|\bs_n\|^2\|\nabla \varphi(\mathbf{\Phi}_{\!n})\|_F}\!>\!\eta_{\mathrm{bfgs}}\\
&\bB_{n}  \quad\;\;\; \mathrm{otherwise}
\end{aligned}
\right.
\end{align}
where $\eta_{\mathrm{bfgs}}\!=\!10^{-6}$ is a small constant. The update rule in \eqref{BFGS4} guarantees that $\bB_n$ is a positive definite matrix in each iteration, and thus $\bS_n$ should be a descent direction. However, due to the roundoff error, sometimes the direction generated by \eqref{BFGS2} may be not a descent direction. To address this numerical issue, we choose $\bS_n$ as
\begin{align}\label{BFGS5}
\mathrm{vec}(\bS_{n})\!=\!\left\{
\begin{aligned}
&\!-\!\bB_{n}\mathrm{vec}[\nabla \varphi(\mathbf{\Phi}_{\!n})] \quad\quad \mathrm{if}\;\xi_n\!>\!\delta_{\mathrm{bfgs}}\\
&\!-\!\mathrm{vec}[\nabla \varphi(\mathbf{\Phi}_{\!n})]  \quad\quad\quad\; \mathrm{otherwise}
\end{aligned}
\right.
\end{align}
where $\xi_n\!=\!\mathrm{vec}[\nabla \varphi(\mathbf{\Phi}_{\!n})]^T\bB_{n}\mathrm{vec}[\nabla \varphi(\mathbf{\Phi}_{\!n})]$ and $\delta_{\mathrm{bfgs}}\!=\!10^{-6}$ is a small constant.

After obtaining the descent direction $\bS_n$, we need to determine the stepsize $\rho_n$ such that the objective function is decreasing in each iteration. We propose a modified backtracking line search method, which is usually more efficient than the classic backtracking line search \cite{boyd2004convex}. The main idea is to use $\rho_{n-1}$ as the initial guess of $\rho_{n}$, and then either increases or decreases it to find the largest $\rho_{n}\!\in\!\mathcal{G}_{n}$ such that
\begin{align}\label{BFGS6}
\mathcal{G}_{n}\!=\!\left\{\rho\!\geq\!0\Bigg|
\begin{aligned}
&\varphi\big(\mathbf{\Phi}_{\!n}\!+\!\rho\bS_n\big)\!\leq\! \varphi(\mathbf{\Phi}_{\!n})+\\
&\rho\!\cdot\!\beta_{\mathrm{bfgs}}\trace\big[\nabla \varphi(\mathbf{\Phi}_{\!n})^T\bS_n\big]
\end{aligned}
\right\}
\end{align}
where $\beta_{\mathrm{bfgs}}\!\in\![0,0.5]$ is a constant to control the stepsize.  Specifically, the stepsize $\rho_n$ is set as
\begin{align}\label{BFGS7}
\rho_{n}\!=\!\left\{
\begin{aligned}
&2^{K_1-1}\!\cdot\!\rho_{n-1} \quad\quad \mathrm{if}\; \rho_{n-1}\!\in\!\mathcal{G}_n\\
&\!\Big(\frac{1}{2}\Big)^{K_2}\!\!\cdot\!\rho_{n-1} \quad\quad \mathrm{if}\; \rho_{n-1}\!\not\in\!\mathcal{G}_n
\end{aligned}
\right.
\end{align}
where $K_1\!\geq\!0$ is the smallest integer such that $2^{K_1}\!\cdot\!\rho_{n-1}\!\not\in\!\mathcal{G}_n$, and $K_2\!\geq\!0$ is the smallest integer such that $(\frac{1}{2})^{K_2}\!\!\cdot\!\rho_{n-1}\!\in\!\mathcal{G}_n$.The details of our BFGS-based algorithm is summarized in Algorithm 1.
\begin{algorithm}
\caption{BFGS-based algorithm}
\begin{algorithmic}
\STATE 1. Inputs: $\bF_{\!\mathrm{opt}}$, $\mathbf{\Phi}_{\!0}$ and $\bB_0$. Set $\rho_0\!=\!1$, $\beta_{\mathrm{bfgs}}\!=\!0.5$, and $\epsilon\!=\!10^{-4}$.
\STATE 2. For $n=0,1,2,...$ (outer iterations)
\begin{itemize}
  \item Determine the descent direction $\bS_n$ by \eqref{BFGS5}.
  \item Compute the stepsize $\rho_n$ via \eqref{BFGS7}.
  \item Update $\mathbf{\Phi}_{\!n}$ to $\mathbf{\Phi}_{\!n+1}$ according to \eqref{BFGS1}.
  \item If $\min\big\{\big|\frac{\varphi(\mathbf{\Phi}_{\!n+1})-\varphi(\mathbf{\Phi}_{\!n})}{\varphi(\mathbf{\Phi}_{\!n+1})}\big|,\|\nabla\varphi(\mathbf{\Phi}_{\!n+1})\|_F\big\}\!<\!\epsilon$, stop.
  \item Update $\bB_n$ to $\bB_{n+1}$ by \eqref{BFGS4}.
\end{itemize}
\STATE 3. Outputs: $\bF_{\!\textrm{RF}}\!=\!\frac{1}{\sqrt{N_\mathrm{T}}}e^{j[\mathbf{0};\mathbf{\Phi}_{\!n}]}$, $\bF_{\!\textrm{BB}}\!=\!\bF_{\!\textrm{RF}}^+\bF_{\!\mathrm{opt}}$.
\end{algorithmic}
\end{algorithm}

According to \cite{li2001global}, the BFGS-based algorithm proposed in Algorithm 1 can converge to a stationary point of problem \eqref{P5}, i.e., the limit of $\nabla \varphi(\mathbf{\Phi}_{\!n})$ satisfies
\begin{align}
\lim_{n\rightarrow \infty}\|\nabla \varphi(\mathbf{\Phi}_{\!n})\|_F\!=\!0.
\end{align}

The performance and convergence speed of Algorithm 1 depends on $\mathbf{\Phi}_{\!0}$ and $\bB_0$. Here a good choice for the initial analog precoder phase is $\angle \bU_\textrm{F}$, where $\angle\bU_\textrm{F}$ is the phase of $\bU_\textrm{F}$. Then the corresponding $\mathbf{\Phi}_{\!0}$ is set as
\begin{align}\label{BFGS8}
\mathbf{\Phi}_{\!0}\!=\!\big[\angle\bU_\textrm{F}\big]_{2:N_\mathrm{T},\bullet}\!-\!\mathbf{1}\big[\angle\bU_\textrm{F}\big]_{1,\bullet}.
\end{align}

The initial inverse Hessian approximation $\bB_0$ will greatly affect the efficiency of Algorithm 1, thus we need to design it carefully. Let the eigendecomposition of $\nabla^2 \varphi(\mathbf{\Phi}_{\!0})$ be
\begin{align}
\nabla^2 \varphi(\mathbf{\Phi}_{\!0})\!=\!\bU_0\bm{\Sigma}_0\bU_0^T
\end{align}
where $\bU_{0}\!\in\!\mathds{C}^{(N_\mathrm{t}-1)N_\mathrm{rf}\times (N_\mathrm{t}-1)N_\mathrm{rf}}$ is a unitary matrix, and $\bm{\Sigma}_0\!\in\!\mathds{R}^{(N_\mathrm{t}-1)N_\mathrm{rf}\times (N_\mathrm{t}-1)N_\mathrm{rf}}$ is a diagonal matrix with eigenvalues arranged in decreasing order. Then $\bB_0$ is given by
\begin{align}\label{BFGS9}
\bB_0\!=\!\bU_0\bm{\hat{\Sigma}}_0^{-1}\bU_0^T
\end{align}
where $\bm{\hat{\Sigma}}_0$ is a diagonal matrix with the $k$-th diagonal entry being
\begin{align}
[\bm{\hat{\Sigma}}_0]_{k,k}\!=\!\left\{
\begin{aligned}
&\big|[\bm{\Sigma}_0]_{k,k}\big| \quad\quad \mathrm{if}\; \big|[\bm{\Sigma}_0]_{k,k}\big|\!\geq\!\delta_{\mathrm{min}}\\
&\;\;\delta_{\mathrm{min}} \quad\quad\quad\; \mathrm{otherwise}.
\end{aligned}
\right.
\end{align}
Here the small constant $\delta_{\mathrm{min}}$ is set as $\delta_{\mathrm{min}}\!=\!10^{-4}$. Since $\bm{\hat{\Sigma}}_0^{-1}$ is a diagonal matrix with positive diagonal entries, the positive definiteness condition of $\bB_0$ is satisfied.

\subsection{Complexity Analysis}
In this subsection, we discuss the per-iteration complexity of the proposed BFGS-based algorithm. Typically, the most time consuming operation in Algorithm 1 is evaluating $\varphi(\mathbf{\Phi})$ and $\nabla\varphi(\mathbf{\Phi})$. Therefore, it is important to analyze the complexity for $\varphi(\mathbf{\Phi})$ and $\nabla\varphi(\mathbf{\Phi})$. Given $\mathbf{\Phi}$, we construct the corresponding analog precoder phase $\bm{\Phi}_\textrm{RF}$ satisfying \eqref{P61} and the analog precoder $\bF_{\!\textrm{RF}}\!=\!\frac{1}{\sqrt{N_\mathrm{t}}}e^{j\bm{\Phi}_\textrm{RF}}$. Then we decompose $\bF_{\!\textrm{RF}}$ by QR decomposition
\begin{align}
\bF_{\!\textrm{RF}}\!=\!\bQ_{\textrm{RF}}\bR_{\textrm{RF}}
\end{align}
where $\bQ_{\textrm{RF}}\!\in\!\mathds{C}^{N_\mathrm{t}\times N_\mathrm{rf}}$ is a unitary matrix, and $\bR_{\textrm{RF}}\!\in\!\mathds{C}^{N_\mathrm{rf}\times N_\mathrm{rf}}$ is an invertible upper triangle matrix. In this way, we can compute $\varphi(\mathbf{\Phi})$ efficiently as
\begin{align}
\varphi(\mathbf{\Phi})\!=\!\|\bF_{\!\mathrm{opt}}\|_F^2\!-\!\|\bQ_{\textrm{RF}}^H\bF_{\!\mathrm{opt}}\|_F^2.
\end{align}
The QR decomposition requires $\mathcal{O}(N_\mathrm{t}N_\mathrm{rf}^2)$ flops, and computing $\|\bQ_{\textrm{RF}}^H\bF_{\!\mathrm{opt}}\|_F^2$ requires $\mathcal{O}(N_\mathrm{t}N_\mathrm{rf}N_\mathrm{s})$ flops. Therefore, the complexity for computing $\varphi(\mathbf{\Phi})$ is about $\mathcal{O}(N_\mathrm{t}N_\mathrm{rf}^2+N_\mathrm{t}N_\mathrm{rf}N_\mathrm{s})$.

The gradient matrix $\nabla \varphi(\mathbf{\Phi})$ can be expressed as
\begin{align}
\nabla \varphi(\mathbf{\Phi})\!=\!\big[\nabla \psi(\bm{\Phi}_\textrm{RF})\big]_{2:N_\mathrm{t},\bullet}
\end{align}
where $\nabla \psi(\bm{\Phi}_\textrm{RF})$ can be expressed using QR decomposition
\begin{align}
\nabla \psi(\bm{\Phi}_\textrm{RF})\!=\!2\Im\big[(\bQ_{\textrm{RF}}\bZ_\textrm{RF}\!-\!\bF_{\!\mathrm{opt}})\bZ_\textrm{RF}^H(\bR_{\textrm{RF}}^{\!-1})^H\!\circ\! \bF_{\!\textrm{RF}}^*\big].
\end{align}
Here $\bZ_\textrm{RF}\!=\!\bQ_{\textrm{RF}}^H\bF_{\!\mathrm{opt}}$. Then the complexity for computing $\nabla \varphi(\mathbf{\Phi})$ is about $\mathcal{O}(N_\mathrm{t}N_\mathrm{rf}^2\!+\!N_\mathrm{rf}^3\!+\!N_\mathrm{s}N_\mathrm{rf}^2\!+\!N_\mathrm{t}N_\mathrm{rf}N_\mathrm{s})$.

Finally, since $\bB_n\!\in\!\mathds{R}^{(N_\mathrm{t}-1)N_\mathrm{rf}\times (N_\mathrm{t}-1)N_\mathrm{rf}}$, the updating rule in \eqref{BFGS4} requires $\mathcal{O}([N_\mathrm{t}-1]^2N_\mathrm{rf}^2)$ flops. Then the per-iteration complexity of Algorithm 1 is given by
\begin{align}
\mathcal{O}\big(N_\mathrm{t}N_\mathrm{rf}^2\!+\!N_\mathrm{rf}^3\!+\!N_\mathrm{s}N_\mathrm{rf}^2\!+\!N_\mathrm{t}N_\mathrm{rf}N_\mathrm{s}\!+\![N_\mathrm{t}-1]^2N_\mathrm{rf}^2\big).
\end{align}

\section{Simulation Results}
\subsection{Average Euclidean Error Evaluation}
The proposed BFGS-based algorithm solves a general constant modulus matrix factorization problem
\begin{equation}\label{CMMF}
\begin{aligned}
& \underset{\bF_{\!\textrm{RF}}\in \mathcal{U},\bF_{\!\textrm{BB}}}{\mathrm{minimize}}
& & \|\bF_{\!\mathrm{opt}}\!-\!\bF_{\!\textrm{RF}}\bF_{\!\textrm{BB}}\|_{F}^2.
\end{aligned}
\end{equation}
Therefore, it is of interest to evaluate the performance of our proposed algorithm for arbitrary given matrix $\bF_{\!\mathrm{opt}}$.

\begin{figure}[h]
  \begin{center}
  \includegraphics[scale=0.55]{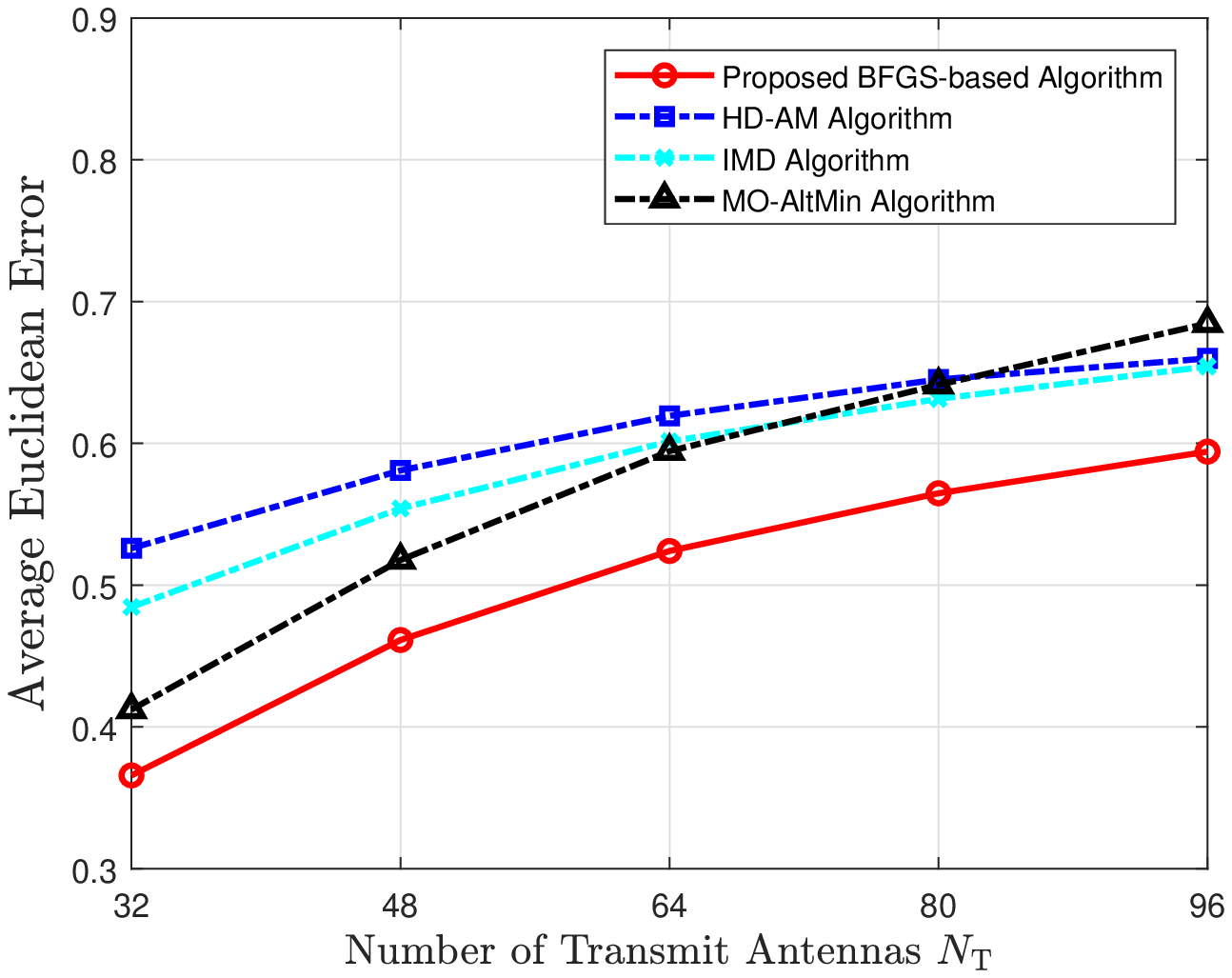}
  \vspace{-0.1cm}
  \caption{Average Euclidean error versus $N_\mathrm{t}$ with 500 randomly generated full rank $\bF_{\!\mathrm{opt}}$.}
\end{center}
  \vspace{-0.6cm}
\end{figure}

We generate $N$ independent samples $\bF_{\!\mathrm{opt}}^{(i)}\!\in\!\mathds{C}^{N_\mathrm{t}\times N_\mathrm{s}}$, $i\!=\!1,2,...,N$ with i.i.d. zero-mean unit-variance complex Gaussian entries. Each sample is then normalized to satisfy
\begin{align}
\big\|\bF_{\!\mathrm{opt}}^{(i)}\big\|_F^2\!=\!N_\mathrm{s},\;\;i=1,2,...,N.
\end{align}
Subsequently, we evaluate the performance of our proposed algorithm by the average Euclidean error, given by
\begin{align}\label{SR1}
\frac{1}{N}\sum_{i=1}^{N}\big\|\bF_{\!\mathrm{opt}}^{(i)}\!-\!\bF_{\!\textrm{RF}}^{(i)}\bF_{\!\textrm{BB}}^{(i)}\big\|_F^2
\end{align}
where $\bF_{\!\textrm{RF}}^{(i)}\!\in\!\mathds{C}^{N_\mathrm{t}\times N_\mathrm{rf}}$ and $\bF_{\!\textrm{BB}}^{(i)}\!\in\!\mathds{C}^{N_\mathrm{rf}\times N_\mathrm{s}}$ are outputs of Algorithm 1 with the given input $\bF_{\!\mathrm{opt}}^{(i)}$.

We make head-to-head comparisons between our proposed BFGS-based algorithm and three existing algorithms, namely the manifold optimization based alternating minimization (MO--AltMin)\cite{yu2016alternating}, the iterative matrix decomposition (IMD) \cite{rajashekar2017iterative} and the hybrid design by alternating minimization (HD--AM) \cite{rusu2016low}. To the best of our knowledge, these three algorithms are the best existing algorithms based on the matrix factorization approach.
Note that the authors in \cite{yu2016alternating} and \cite{rajashekar2017iterative} claim that their proposed algorithms have significant performance gains over other existing algorithms, and the authors in \cite{rusu2016low} claim that the HD--AM algorithm provides the best solution among four different hybrid precoding algorithms proposed in \cite{rusu2016low}. Therefore, if the proposed BFGS-based algorithm can beat these algorithms, we believe it outperforms other existing algorithms based on the matrix factorization approach.

The matrix factorization based algorithms \cite{yu2016alternating,rusu2016low,rajashekar2017iterative} involve a normalization procedure to ensure $\|\bF_{\!\textrm{RF}}\bF_{\!\textrm{BB}}\|_F^2\!=\!P$. Since the mutual information is monotonically increasing with respect to $\|\bF_{\!\textrm{RF}}\bF_{\!\textrm{BB}}\|_F$, this procedure will increase the achievable rate. However, when we choose the Euclidean error as the performance metric, the normalization procedure will decrease the overall performance because these algorithms and our proposed BFGS-based algorithm are designed to solve problem \eqref{CMMF} without the equality power constraint. Therefore, for the sake of fairness, we do not execute the normalization for all algorithms in this subsection.

\begin{table*}[t]
\centering
\begin{tabular}{l*{6}{c}r}
$N_\mathrm{T}$         & 32 & 48 & 64 & 80 & 96  \\
\hline
Proposed BFGS-based algorithm     & 0.014s & 0.021s & 0.034s & 0.033s & 0.149s \\
HD--AM algorithm                  & 0.008s & 0.009s & 0.014s & 0.017s & 0.022s \\
IMD algorithm                     & 0.012s & 0.013s & 0.019s & 0.020s & 0.022s                               \\
MO--AltMin algorithm              & 0.349s & 0.696s & 1.226s & 1.924s & 5.429s \\
\end{tabular}
\vspace{-0.1cm}
\caption{Average running time (in secs.) versus $N_\mathrm{T}$ with 500 randomly generated full rank $\bF_{\!\mathrm{opt}}$.}
\vspace{-0.1cm}
\end{table*}
\begin{table*}[ht]
\centering
\begin{tabular}{l*{6}{c}r}
$\mathrm{SNR}(\mathrm{dB})$       & -35 & -30 & -25 & -20 & -15 & -10 & -5 \\
\hline
WF algorithm (benchmark)          &0.0767&0.2276&0.6169&1.4544&3.0283&5.5588&9.2619\\
Proposed BFGS-based algorithm     &0.0763&0.2265&0.6141&1.4515&3.0175&5.4993&9.1137\\
HD--AM algorithm                  &0.0696&0.2080&0.5685&1.3614&2.8641&5.3070&8.9276\\
IMD algorithm                     &0.0698&0.2086&0.5702&1.3649&2.8707&5.3164&8.9382\\
MO--AltMin algorithm              &0.0743&0.2189&0.6033&1.4375&2.9878&5.4584&9.0379 \\
\end{tabular}
\vspace{-0.1cm}
\caption{Average mutual information with Gaussian inputs versus $\mathrm{SNR}$ for various algorithms.}
\vspace{-0.4cm}
\end{table*}

We set the number of samples as $N\!=\!500$, and $N_\mathrm{rf}$ and $N_\mathrm{s}$ are restricted to be $N_\mathrm{rf}\!=\!N_\mathrm{s}\!=\!4$. The initial analog precoders for these four algorithms are set as $\angle\bF_{\!\mathrm{opt}}^{(i)}$. The average Euclidean error and average running time of four algorithms are presented in
Fig. 1 and Table I. From Fig. 1 and Table I, we have the following remarks:
\begin{enumerate}
  \item The proposed BFGS-based algorithm and the MO-AltMin algorithm are guaranteed to converge to the stationary point of problem \eqref{CMMF}, while the HD--AM and IMD algorithms may not achieve this goal.
  \item The proposed BFGS-based algorithm significantly outperforms the HD--AM, IMD and MO--AltMin algorithms in the whole range of $N_\mathrm{T}$. In addition, it consumes much lower computational time than the MO--AltMin algorithm.
  \item The phenomenon that the BFGS-based algorithm outperforms the MO-AltMin algorithm can be explained as follows. For nonconvex problem \eqref{CMMF}, its stationary points can be local minimum (positive definite Hessian), local maximum (negative definite Hessian), or saddle point (indefinite Hessian). Most stationary points are saddle points in high dimensional space, and the objective value at the saddle point is usually worse than that at the local optimum \cite{dauphin2014identifying}. In order to decrease the possibility for converging to the saddle point, we can 1) decrease the dimensions of the search space; 2) use Hessian information to avoid converging to the indefinite Hessian point\cite{dauphin2014identifying,reddi2017generic}. Since the proposed BFGS-based algorithm utilize these two techniques to avoid saddle points, its performance is better than that of the MO-AltMin algorithm.
\end{enumerate}

\subsection{Average Mutual Information Evaluation With Gaussian Inputs}
We consider a $4\times 72$ MIMO system with $N_\mathrm{rf}\!=\!N_\mathrm{s}\!=\!4$. The number of physical propagation paths is set as $L\!=\!8$, and the signal-to-noise ratio (SNR) is defined as $\mathrm{SNR}\!=\!\frac{P}{\sigma^2}$. We generate $N\!=\!1000$ channel realizations by \eqref{MCM}, and evaluate the system performance by the following average mutual information with Gaussian inputs:
\begin{align}
\frac{1}{N}\sum_{i=1}^N\log\det\Big[\bI+\sigma^{-2}\bH_i\bQ_i\bH_i^H\Big]
\end{align}
where $\bH_i$ is the $i$th channel realization, and $\bQ_i\!=\!\bF_{\!\textrm{RF}}^{(i)}\bF_{\!\textrm{BB}}^{(i)}(\bF_{\!\textrm{BB}}^{(i)})^H(\bF_{\!\textrm{RF}}^{(i)})^H$ with $(\bF_{\!\textrm{RF}}^{(i)},\bF_{\!\textrm{BB}}^{(i)})$ being the analog and digital precoder solution corresponding to $\bH_i$.

We set the performance of unconstrained optimal precoder as a benchmark, and then compare our proposed BFGS-based algorithm with the IMD algorithm, the HD--AM algorithm and the MO--AltMin algorithm. The unconstrained optimal precoder $\bF_{\!\mathrm{opt}}$ under Gaussian inputs can be obtained by the waterfilling (WF) algorithm, and all hybrid precoding algorithms in this subsection use the same $\bF_{\!\mathrm{opt}}$ to design analog and digital precoders. Moreover, the initial analog precoders of these algorithms are set as $\bF_{\!\textrm{RF}}\!=\!\frac{1}{\sqrt{N_\mathrm{t}}}e^{j[\bV_{\!\textrm{H}}]_{\bullet,1:N_\mathrm{rf}}}$, where $[\bV_{\!\textrm{H}}]_{\bullet,1:N_\mathrm{rf}}$ is the first $N_\mathrm{rf}$ right singular vectors  of $\bH$.

Table II demonstrates the average mutual information with Gaussian inputs versus SNR for various algorithms. From Table II, we have the following remarks:
\begin{enumerate}
  \item The proposed BFGS-based algorithm has about 10$\%$ performance gain over HD--AM and IMD algorithms in low SNR regimes because HD--AM and IMD algorithms are designed for full rank $\bF_{\!\mathrm{opt}}$. However, the unconstrained optimal precoder $\bF_{\!\mathrm{opt}}$ is not a full rank matrix in low SNR regimes. In addition, the HD--AM and IMD algorithms can be applied only when $N_\mathrm{rf}\!=\!N_{\mathrm{s}}$, while our proposed BFGS-based algorithm and the MO-AltMin algorithm can work for arbitrary $N_\mathrm{rf}$ and $N_{\mathrm{s}}$.
  \item When the unconstrained optimal precoder is obtained by WF algorithm and the performance metric is chosen as the average mutual information, the gain of our proposed BFGS-based algorithm over the MO-AltMin algorithm is not very significant compared with Fig. 1. However, as shown in Table I, our proposed BFGS-based algorithm is much faster than the MO-AltMin algorithm. Therefore, our proposed BFGS-based algorithm also has advantages over the MO-AltMin algorithm.
\end{enumerate}

\subsection{Average Mutual Information Evaluation With Finite-Alphabet Inputs}
We first consider a $64\times 64$ MIMO system with $N_\mathrm{rf}\!=\!N_{\mathrm{s}}\!=\!4$. The number of physical propagation paths is set as $L\!=\!6$. The input signal is drawn from QPSK modulation, and SNR is defined as $\mathrm{SNR}\!=\!\frac{P}{\sigma^2}$. In addition, the system performance is measured by the average mutual information, which is averaged over 1000 channel realizations generated by \eqref{MCM}.
\begin{figure}[h]
  \begin{center}
  \includegraphics[scale=0.55]{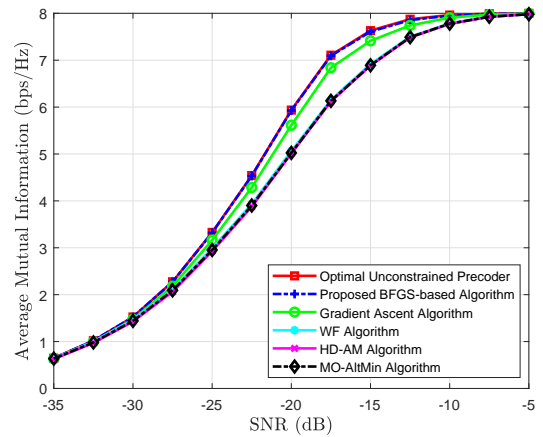}
  \vspace{-0.1cm}
  \caption{Average mutual information versus SNR for different algorithms in a $64\times 64$ system with $N_\mathrm{rf}\!=\!N_{\mathrm{s}}\!=\!4$.}
  \end{center}
  \vspace{-0.6cm}
\end{figure}

We set the unconstrained optimal precoder under finite-alphabet inputs as a benchmark, and then compare our proposed BFGS-based algorithm with the gradient ascent algorithm \cite{rajashekar2016hybrid}, the classic waterfilling (WF) algorithm, the HD--AM algorithm\cite{rusu2016low} and the MO--AltMin algorithm\cite{yu2016alternating}. For fair comparisons, the initial analog precoders of these algorithms are set as $\bF_{\!\textrm{RF}}\!=\!\frac{1}{\sqrt{N_\mathrm{t}}}e^{j[\bV_{\!\textrm{H}}]_{\bullet,1:N_\mathrm{rf}}}$.

Among these algorithms, our proposed BFGS-based algorithm and the gradient ascent algorithm are designed for finite-alphabet inputs, and the remaining three algorithms are designed under Gaussian inputs. Specifically, the HD--AM and MO--AltMin algorithms decompose the WF optimal precoder into digital and analog precoders, and then evaluate the corresponding mutual information under finite-alphabet inputs.

Fig. 2 demonstrates the average mutual information versus SNR for different algorithms. The results in Fig. 2 imply three observations. First, our proposed BFGS-based algorithm has the potential to achieve the performance of unconstrained optimal precoders. Second, our algorithm has about 0.2 bps/Hz improvement compared to the gradient ascent algorithm. Since mmWave provide very large bandwidths, a gain of 0.2 bps/Hz would translate to a large increase in the effective data rate. Third, the proposed BFGS-based algorithm has about 3dB gain over the HD--AM and MO--AltMin algorithms. This is mainly because the unconstrained optimal precoder designed under Gaussian inputs will lead to significant performance loss when applying to finite-alphabet systems.

Next, we consider a $32\times 80$ MIMO system with $L\!=\! 8$, $N_\mathrm{rf}\!=\!6$ and $N_{\mathrm{s}}\!=\!4$. The input signal is drawn from QPSK modulation. In this case, the gradient ascent and HD--AM algorithms cannot work because they assume $N_{\mathrm{s}}\!=\!N_\mathrm{rf}$.
\begin{figure}[h]
  \begin{center}
  \includegraphics[scale=0.55]{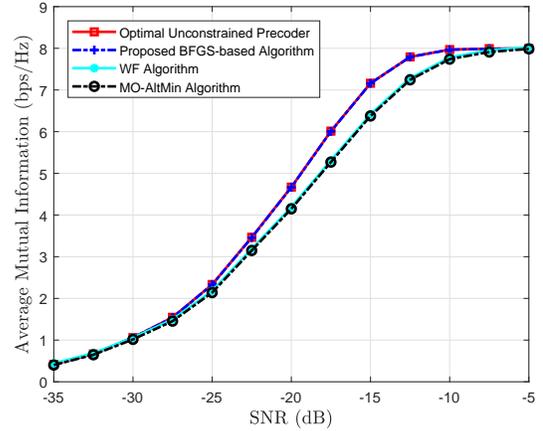}
  \vspace{-0.1cm}
  \caption{Average mutual information versus SNR for different methods in a $32\times80$ system with $N_\mathrm{rf}\!=\!6$ and $N_{\mathrm{s}}\!=\!4$.}
  \end{center}
  \vspace{-0.6cm}
\end{figure}
Therefore, we only compare our proposed BFGS-based algorithm with the MO-AltMin Algorithm. The simulation result is shown in Fig. 3. Based on the results in Fig. 3, we have the following remarks:
\begin{itemize}
  \item The proposed BFGS-based algorithm and the MO--AltMin Algorithm are more general than the gradient ascent and HD--AM algorithms because they can work when $N_{\mathrm{s}}\!<\!N_\mathrm{rf}$.
  \item Our proposed algorithm can achieve the performance of unconstrained optimal precoder in whole SNR regimes. In addition, the MO--AltMin algorithm with WF optimal precoder has about 2--3dB performance loss compared with the our proposed BFGS-based algorithm.
\end{itemize}

\section{Conclusion}
This paper considers the hybrid precoding design for mmWave MIMO systems with finite-alphabet inputs. The precoding problem has been formulated as a matrix factorization problem with constant modulus constraints. We first proposed a sufficient and a necessary condition for the hybrid precoding scheme to achieve the performance of unconstrained optimal precoders. Next, we decoupled the constant modulus matrix factorization problem by showing that the power constraint can be removed without loss of local and/or global optimality. Then we proposed a BFGS-based method to solve the constant modulus matrix factorization problem. Numerical results have demonstrated the effectiveness of our proposed algorithm for hybrid precoding designs in mmWave MIMO systems.
\newpage
\section*{Appendix A\\ Proofs of Propositions 1--3 and Theorem 1}
\begin{IEEEproof}[Proof of Proposition 1]
If there exists a full rank square matrix $\bS$ such that $\bU_{\textrm{F}}\bS\!\in\! \mathcal{U}$, we can construct $\bF_{\!\textrm{RF}}$ and $\bF_{\!\textrm{BB}}$ as
\begin{align}
\bF_{\!\textrm{RF}}\!=\!\bU_{\textrm{F}}\bS,\;\bF_{\!\textrm{BB}}\!=\!\bS^{-1}\mathbf{\Sigma}_{\textrm{F}}\bV_{\!\textrm{F}}^H
\end{align}
where $\mathbf{\Sigma}_{\textrm{F}}$ is a diagonal matrix with singular values of $\bF_{\!\mathrm{opt}}$ arranged in decreasing order, and $\bV_{\!\textrm{F}}$ is a unitary matrix with right singular vectors of $\bF_{\!\mathrm{opt}}$. Then
\begin{align}
\bF_{\!\textrm{RF}}\!\in\!\mathcal{U},\;\bF_{\!\textrm{RF}}\bF_{\!\textrm{BB}}\!=\!\bU_{\textrm{F}}\mathbf{\Sigma}_{\textrm{F}}\bV_{\!\textrm{F}}^H\!=\!\bF_{\!\mathrm{opt}}.
\end{align}

Conversely, if there exists $(\bF_{\!\textrm{RF}},\bF_{\!\textrm{BB}})$ such that $\bF_{\!\mathrm{opt}}\!=\!\bF_{\!\textrm{RF}}\bF_{\!\textrm{BB}}$, $\mathcal{C}(\bF_{\!\mathrm{opt}})$ is a subspace of $\mathcal{C}(\bF_{\!\textrm{RF}})$, where $\mathcal{C}(\cdot)$ represents the space spanned by columns of a matrix. Moreover, according to
$\bF_{\!\mathrm{opt}}\!=\!\bU_{\textrm{F}}\mathbf{\Sigma}_{\textrm{F}}\bV_{\!\textrm{F}}^H$, the first $\mathrm{rank}(\bF_{\!\mathrm{opt}})$ columns of $\bU_\textrm{F}$ form an orthogonal basis of $\mathcal{C}(\bF_{\!\mathrm{opt}})$. Since $\mathcal{C}(\bF_{\!\mathrm{opt}})$ is a subspace of $\mathcal{C}(\bF_{\!\textrm{RF}})$, we can use the Gram-Schmidt algorithm to construct the remaining $N_\mathrm{rf}-\mathrm{rank}(\bF_{\!\mathrm{opt}})$ columns of $\bU_\textrm{F}$ such that the columns of $\bU_\textrm{F}$ form an orthogonal basis of $\mathcal{C}(\bF_{\!\textrm{RF}})$. Then there exists a full rank matrix $\bS$ satisfying $\bF_{\!\textrm{RF}}\!=\!\bU_{\textrm{F}}\bS\!\in\! \mathcal{U}$. This completes the proof.
\end{IEEEproof}

\begin{IEEEproof}[Proof of Proposition 2]
Let the SVD of $\bH$ be
\begin{align}
\bH\!=\!\bU_{\textrm{H}}\mathbf{\Sigma}_{\textrm{H}}\bV_{\!\textrm{H}}^H
\end{align}
where $\bU_{\textrm{H}}\!\in\!\mathds{C}^{N_{\mathrm{r}}\times \mathrm{rank}(\bH)}$ is a unitary matrix with left singular vectors, $\mathbf{\Sigma}_{\textrm{H}}\!\in\!\mathds{C}^{\mathrm{rank}(\bH)\times \mathrm{rank}(\bH)}$ is a diagonal matrix with singular values arranged in decreasing order, and $\bV_{\!\textrm{H}}\!\in\!\mathds{C}^{N_{\mathrm{t}}\times \mathrm{rank}(\bH)}$ is a unitary matrix with right singular vectors. Based on equation \eqref{rank}, when $L\!\leq \min(N_{\mathrm{r}},N_{\mathrm{t}})$, $\mathrm{rank}(\bH)\!=\!L$. Then the columns of $\bV_{\!\textrm{H}}$ form an orthogonal basis of $\mathcal{C}(\bH^H)$. Moreover, since $\bH\!=\!\bA_\mathrm{r}\diag(\bm{\alpha})\bA_\mathrm{t}^{\!H}$ and $\mathrm{rank}(\bA_\mathrm{t})\!=\!L$, the columns of $\bA_\mathrm{t}$ also form a basis of $\mathcal{C}(\bH^H)$. Therefore, there exists a full rank square matrix $\bS\!\in\!\mathds{C}^{L\times L}$ such that $\bA_\mathrm{t}\!=\!\bV_{\!\textrm{H}}\bS\!\in\!\mathcal{U}$. The semi-unitary matrix $\bV_{\!\textrm{H}}$ has a close connection with the left singular vectors of $\bF_{\!\mathrm{opt}}$. Specifically, the left singular vectors of $\bF_{\!\mathrm{opt}}$ can always be chosen as the first $N_{\mathrm{s}}$ columns of $\bV_{\!\mathrm{H}}$ \cite[Proposition 2]{xiao2011globally}, i.e.,
\begin{align}
\bU_\textrm{F}\!=\!\big[{\bV_{\!\mathrm{H}}}\big]_{\bullet,1:N_\mathrm{s}}.
\end{align}
Therefore, when $L\!=\!N_\mathrm{s}\!=\!\mathrm{min}\{L,N_\mathrm{rf}\}\!\leq\!\mathrm{min}\{N_\mathrm{r},N_\mathrm{t}\}$, we have $\bA_\mathrm{t}\!=\!\tilde{\bV}_{\!\textrm{H}}\bS\!=\!\bU_\textrm{F}\bS\!\in\!\mathcal{U}$. Finally, $L\!=\!\mathrm{min}\{L,N_\mathrm{rf}\}\!\leq\!\mathrm{min}\{N_\mathrm{r},N_\mathrm{t}\}$
holds if and only if $L\!\leq\!\mathrm{min}\{N_\mathrm{r},N_\mathrm{t},N_\mathrm{rf}\}$. This completes the proof.
\end{IEEEproof}

\begin{IEEEproof}[Proof of Proposition 3]
We first rewrite the solutions of $\bK_\mathrm{F}\mathrm{vec}(\bZ)\!=\!\mathbf{1}$ as
\begin{align}
\mathrm{vec}(\bZ)\!=\!\bm{\xi}_0+\sum_{i=1}^I\alpha_i\bm{\xi}_i.
\end{align}
Here $\bm{\xi}_0$ is a particular solution to $\bK_\mathrm{F}\mathrm{vec}(\bZ)\!=\!\mathbf{1}$, $\{\alpha_i\}_{i=1}^I$ are complex numbers, and $\{\bm{\xi}_i\}_{i=1}^I$ is a basis of $\mathcal{N}(\bK_\mathrm{F})$, where $\mathcal{N}(\cdot)$ represents the null space of a matrix. Since the nonliear equations
\begin{equation}
\begin{aligned}
\bK_\mathrm{F}\mathrm{vec}(\bZ)\!=\!\mathbf{1},\;\bZ\!\succeq\!\bm{0},\;\mathrm{rank}(\bZ)\!=\!1
\end{aligned}
\end{equation}
have $N_{\mathrm{RF}}$ linear independent solutions, the dimension of $\mathcal{N}(\bK_\mathrm{F})$ should be at least $N_{\mathrm{RF}}-1$, which implies
\begin{align}
\mathrm{dim}\big[\mathcal{N}(\bK_\mathrm{F})\big]\!=\!N_{\mathrm{RF}}^2\!-\!\mathrm{rank}(\bK_\mathrm{F})\!\geq\! N_{\mathrm{RF}}\!-\!1.
\end{align}
This completes the proof.
\end{IEEEproof}

\begin{IEEEproof}[Proof of Theorem 1]
If $(\hat{\bF}_{\!\textrm{RF}},\hat{\bF}_{\!\textrm{BB}})$ is a KKT point of problem \eqref{P3}, then it satisfies the following KKT conditions:
\begin{align}
-(\bF_{\!\mathrm{opt}}\!-\!\hat{\bF}_{\!\textrm{RF}}\hat{\bF}_{\!\textrm{BB}})\hat{\bF}_{\!\textrm{BB}}^H+\mathbf{\Upsilon}\circ\hat{\bF}_{\!\textrm{RF}}=\mathbf{0}\\
\hat{\bF}_{\!\textrm{RF}}^H(\bF_{\!\mathrm{opt}}\!-\!\hat{\bF}_{\!\textrm{RF}}\hat{\bF}_{\!\textrm{BB}})=\mathbf{0}\label{T1}\\
\hat{\bF}_{\!\textrm{RF}}^*\circ\hat{\bF}_{\!\textrm{RF}}\!=\! \frac{1}{N_{\mathrm{t}}}\mathbf{1}
\end{align}
where $\mathbf{\Upsilon}_{\!ij}$ is the lagrangian multiplier associated with the equality constraint $[\bF_{\!\textrm{RF}}]_{ij}^*[\bF_{\!\textrm{RF}}]_{ij}\!=\! 1/N_{\mathrm{t}}$. Suppose that $\hat{\bF}_{\!\textrm{RF}}$ has full column rank, then equation \eqref{T1} becomes
\begin{align}\label{T2}
\hat{\bF}_{\!\textrm{BB}}=\hat{\bF}_{\!\textrm{RF}}^{+}\bF_{\!\mathrm{opt}}.
\end{align}
where $\hat{\bF}_{\!\textrm{RF}}^{+}\!=\!(\hat{\bF}_{\!\textrm{RF}}^H\hat{\bF}_{\!\textrm{RF}})^{-1}\hat{\bF}_{\!\textrm{RF}}^H$ is the Moore-Penrose pseudoinverse of $\bF_{\!\textrm{RF}}$.  Inserting equation \eqref{T2} into $\trace\big(\hat{\bF}_{\!\textrm{BB}}^H\hat{\bF}_{\!\textrm{RF}}^H\hat{\bF}_{\!\textrm{RF}}\hat{\bF}_{\!\textrm{BB}}\big)$, we obtain
\begin{align}
\trace&\big(\hat{\bF}_{\!\textrm{BB}}^H\hat{\bF}_{\!\textrm{RF}}^H\hat{\bF}_{\!\textrm{RF}}\hat{\bF}_{\!\textrm{BB}}\big)\nonumber\\
&=\!\trace\big(\bF_{\!\mathrm{opt}}^H\hat{\bF}_{\!\textrm{RF}}\hat{\bF}_{\!\textrm{RF}}^{+}\bF_{\!\mathrm{opt}}\big)\!=\!\trace\big(\hat{\bF}_{\!\textrm{RF}}\hat{\bF}_{\!\textrm{RF}}^{+}\bF_{\!\mathrm{opt}}\bF_{\!\mathrm{opt}}^H)\nonumber\\
&\leq\! \sum_{i=1}^{N_{\mathrm{t}}} \lambda_{i}\big(\hat{\bF}_{\!\textrm{RF}}\hat{\bF}_{\!\textrm{RF}}^{+}\big)\lambda_{i}\big(\bF_{\!\mathrm{opt}}\bF_{\!\mathrm{opt}}^H\big)\label{T3}
\end{align}
where $\lambda_{i}(\cdot)$ represents the eigenvalue of a Hermitian matrix in decreasing order. The inequality in \eqref{T3} follows from \cite[Lemma II.1]{lasserre1995trace}:
\begin{align}
\sum_{i=1}^{n}\lambda_{i}(\bA)\lambda_{n-i+1}(\bB)\!\leq\!\trace(\bA\bB)\!\leq\!\sum_{i=1}^{n}\lambda_{i}(\bA)\lambda_{i}(\bB)
\end{align}
where $\bA\!\in\! \mathds{C}^{n\times n}$ and $\bB\!\in\! \mathds{C}^{n\times n}$ are Hermitian matrices. Since $\hat{\bF}_{\!\textrm{RF}}\hat{\bF}_{\!\textrm{RF}}^+$ is a projection matrix, its eigenvalues are given by
\begin{align}
\lambda_{i}(\hat{\bF}_{\!\textrm{RF}}\hat{\bF}_{\!\textrm{RF}}^+)=\left\{
\begin{aligned}
&1, \quad i=1,2,...,N_\mathrm{rf}\\
&0, \quad \mathrm{otherwise}
\end{aligned}
\right.
\end{align}
Then $\trace\big(\hat{\bF}_{\!\textrm{BB}}^H\hat{\bF}_{\!\textrm{RF}}^H\hat{\bF}_{\!\textrm{RF}}\hat{\bF}_{\!\textrm{BB}}\big)$ can be further upper bounded by
\begin{align*}
&\trace\big(\hat{\bF}_{\!\textrm{BB}}^H\hat{\bF}_{\!\textrm{RF}}^H\hat{\bF}_{\!\textrm{RF}}\hat{\bF}_{\!\textrm{BB}}\big) \!\leq\!\sum_{i=1}^{N_\mathrm{rf}}\lambda_{i}\big(\bF_{\!\mathrm{opt}}\bF_{\!\mathrm{opt}}^H\big)\!\leq\! \trace(\bF_{\!\mathrm{opt}}\bF_{\!\mathrm{opt}}^H)\!=\! P.
\end{align*}
This completes the proof.
\end{IEEEproof}

\section*{Appendix B\\Proofs of Theorem 2--3 and Lemma 1}
\begin{IEEEproof}[Proof of Theorem 2]
The KKT conditions of problem \eqref{P3} are given by
\begin{align}
-(\bF_{\!\mathrm{opt}}\!-\!\bF_{\!\textrm{RF}}\bF_{\!\textrm{BB}})\bF_{\!\textrm{BB}}^{H}\!+\!\mathbf{\Upsilon}\!\circ\!\bF_{\!\textrm{RF}}\!=\!\mathbf{0}\label{B1}\\
\bF_{\!\textrm{RF}}^{H}(\bF_{\!\mathrm{opt}}\!-\!\bF_{\!\textrm{RF}}\bF_{\!\textrm{BB}})=\mathbf{0}\label{B2}\\
\bF_{\!\textrm{RF}}^*\!\circ\!\bF_{\!\textrm{RF}}\!=\! \frac{1}{N_{\mathrm{t}}}\mathbf{1}\label{B3}
\end{align}
where $\mathbf{\Upsilon}_{\!ij}$ is the lagrangian multiplier associated with the equality constraint $[\bF_{\!\textrm{RF}}]_{kl}^*[\bF_{\!\textrm{RF}}]_{kl}\!=\! 1/N_{\mathrm{t}}$. Suppose $\hat{\bF}_{\!\textrm{RF}}$ is a KKT point of problem \eqref{P4} and $\hat{\bF}_{\!\textrm{BB}}\!=\!\hat{\bF}_{\!\textrm{RF}}^{+}\bF_{\!\mathrm{opt}}$, $(\hat{\bF}_{\!\textrm{RF}}, \hat{\bF}_{\!\textrm{BB}})$ satisfies equations \eqref{B2} and \eqref{B3}. Moreover, $\hat{\bF}_{\!\textrm{RF}}$ satisfies the following stationarity condition of problem \eqref{P4}:
\begin{align}\label{B4}
-\big(\bI\!-\!\hat{\bF}_{\!\textrm{RF}}\hat{\bF}_{\!\textrm{RF}}^{+}\big)\bF_{\!\mathrm{opt}}\bF_{\!\mathrm{opt}}^{H}(\hat{\bF}_{\!\textrm{RF}}^{+})^{H}\!+\!\mathbf{\Upsilon}\!\circ\!\hat{\bF}_{\!\textrm{RF}}\!=\! \mathbf{0}
\end{align}
where $-\big(\bI\!-\!\hat{\bF}_{\!\textrm{RF}}\hat{\bF}_{\!\textrm{RF}}^{+}\big)\bF_{\!\mathrm{opt}}\bF_{\!\mathrm{opt}}^{H}(\hat{\bF}_{\!\textrm{RF}}^{+})^{H}$ is the complex gradient of $f(\bF_{\!\textrm{RF}})$, and $\mathbf{\Upsilon}$ is the lagrangian multiplier. Inserting $\hat{\bF}_{\!\textrm{BB}}\!=\!\hat{\bF}_{\!\textrm{RF}}^{+}\bF_{\!\mathrm{opt}}$ into equation \eqref{B4}, it becomes
\begin{align}
-(\hat{\bF}_{\!\mathrm{opt}}\!-\!\hat{\bF}_{\!\textrm{RF}}\hat{\bF}_{\!\textrm{BB}})\hat{\bF}_{\!\textrm{BB}}^{H}\!+\!\mathbf{\Upsilon}\!\circ\!\hat{\bF}_{\!\textrm{RF}}\!=\!\mathbf{0}
\end{align}
which is exactly the stationarity condition of problem \eqref{P3} given in equation \eqref{B1}. Therefore, the KKT point of problem \eqref{P4} satisfies equations \eqref{B1}--\eqref{B3} and it is a KKT point of problem \eqref{P3}.

Suppose that $\hat{\bF}_{\!\textrm{RF}}$ is a globally optimal solution of problem \eqref{P4} and $\hat{\bF}_{\!\textrm{BB}}\!=\!\hat{\bF}_{\!\textrm{RF}}^{+}\bF_{\!\mathrm{opt}}$, then
\begin{align}
r(\hat{\bF}_{\!\textrm{RF}},\hat{\bF}_{\!\textrm{BB}})\!=\!f(\hat{\bF}_{\!\textrm{RF}})
\end{align}
where $r(\bF_{\!\textrm{RF}},\bF_{\!\textrm{BB}})\!=\!\|\bF_{\!\mathrm{opt}}\!-\!\bF_{\!\textrm{RF}}\bF_{\!\textrm{BB}}\|_{F}^2$. We further assume $(\hat{\bF}_{\!\textrm{RF}},\hat{\bF}_{\!\textrm{BB}})$ is not a globally optimal solution of problem \eqref{P3}, i.e., there exists a feasible solution $(\tilde{\bF}_{\!\textrm{RF}}, \tilde{\bF}_{\!\textrm{BB}})$ such that $r(\tilde{\bF}_{\!\textrm{RF}}, \tilde{\bF}_{\!\textrm{BB}})\!<\!r(\hat{\bF}_{\!\textrm{RF}},\hat{\bF}_{\!\textrm{BB}})$. Since for any given $\bF_{\!\textrm{BB}}$, $f(\bF_{\!\textrm{RF}})\!\leq\! r(\bF_{\!\textrm{RF}},\bF_{\!\textrm{BB}})$, we have
\begin{align}
f(\tilde{\bF}_{\!\textrm{RF}})\!\leq \! r(\tilde{\bF}_{\!\textrm{RF}}, \tilde{\bF}_{\!\textrm{BB}})\!<\!r(\hat{\bF}_{\!\textrm{RF}},\hat{\bF}_{\!\textrm{BB}})\!=\! f(\hat{\bF}_{\!\textrm{RF}})
\end{align}
which is a contradiction to the fact that $\hat{\bF}_{\!\textrm{RF}}$ is a globally optimal solution of problem \eqref{P4}. Therefore, $(\hat{\bF}_{\!\textrm{RF}},\hat{\bF}_{\!\textrm{BB}})$ is a globally optimal solution of problem \eqref{P3}.

Conversely, suppose that $(\hat{\bF}_{\!\textrm{RF}},\hat{\bF}_{\!\textrm{BB}})$ is a globally optimal solution of problem \eqref{P3}, then
\begin{align}
r(\hat{\bF}_{\!\textrm{RF}},\hat{\bF}_{\!\textrm{BB}})\!=\!f(\hat{\bF}_{\!\textrm{RF}}).
\end{align}
Similarly, we assume $\hat{\bF}_{\!\textrm{RF}}$ is not a globally optimal solution of problem \eqref{P4}, i.e., there exists a feasible $\tilde{\bF}_{\!\textrm{RF}}$ such that $f(\tilde{\bF}_{\!\textrm{RF}})\!<\!f(\hat{\bF}_{\!\textrm{RF}})$. Let $\tilde{\bF}_{\!\textrm{BB}}\!=\!\tilde{\bF}^+_{\!\textrm{RF}}\bF_{\!\mathrm{opt}}$, then
\begin{align}
f(\tilde{\bF}_{\!\textrm{RF}})\!=\! r(\tilde{\bF}_{\!\textrm{RF}},\tilde{\bF}_{\!\textrm{BB}})\!<\!f(\hat{\bF}_{\!\textrm{RF}})\!=\!r(\hat{\bF}_{\!\textrm{RF}},\hat{\bF}_{\!\textrm{BB}})
\end{align}
which is a contradiction to the fact that $(\hat{\bF}_{\!\textrm{RF}},\hat{\bF}_{\!\textrm{BB}})$ is a globally optimal solution of problem \eqref{P3}. Therefore, $\hat{\bF}_{\!\textrm{RF}}$ is a globally optimal solution of problem \eqref{P4}. This completes the proof.
\end{IEEEproof}

\begin{IEEEproof}[Proof of Lemma 1]
We first compute the complex gradient matrix $\nabla_{\!\bF_{\!\textrm{RF}}} f(\bF_{\!\textrm{RF}})$. Note that $f(\bF_{\!\textrm{RF}})$ can be rewritten as
\begin{align}
f(\bF_{\!\textrm{RF}})\!=\!\|\bF_{\!\mathrm{opt}}\|_{F}^2\!-\!\trace\big(\bF_{\!\textrm{RF}}^{+}\bF_{\!\mathrm{opt}}\bF_{\!\mathrm{opt}}^H\bF_{\!\textrm{RF}}\big).
\end{align}
Then the differential of $f(\bF_{\!\textrm{RF}})$ is given by
\begin{align}\label{C1}
\mathrm{d}f(\bF_{\!\textrm{RF}})\!=\!-\trace\big(\mathrm{d}\bF_{\!\textrm{RF}}^{+}\bF_{\!\mathrm{opt}}\bF_{\!\mathrm{opt}}^H\bF_{\!\textrm{RF}}\big)\!-\!\trace\big(\bF_{\!\textrm{RF}}^{+}\bF_{\!\mathrm{opt}}\bF_{\!\mathrm{opt}}^H\mathrm{d}\bF_{\!\textrm{RF}}\big).
\end{align}
The differential of $\bF_{\!\textrm{RF}}^{+}\!=\!(\bF_{\!\textrm{RF}}^H\bF_{\!\textrm{RF}})^{-1}\bF_{\!\textrm{RF}}^H$ in equation \eqref{C1} can be computed as follows:
\begin{align}\label{C2}
\mathrm{d}\bF_{\!\textrm{RF}}^{+}&\!=\!\mathrm{d}\big[(\bF_{\!\textrm{RF}}^H\bF_{\!\textrm{RF}})^{-1}\big]\bF^H_{\!\textrm{RF}}+(\bF_{\!\textrm{RF}}^H\bF_{\!\textrm{RF}})^{-1}\mathrm{d}\bF_{\!\textrm{RF}}^H\nonumber\\
&\!=\!(\bF_{\!\textrm{RF}}^H\bF_{\!\textrm{RF}})^{-1}\mathrm{d}\bF_{\!\textrm{RF}}^H(\bI-\bF_{\!\textrm{RF}}\bF_{\!\textrm{RF}}^{+})\!-\!\bF^+_{\!\textrm{RF}}\mathrm{d}\bF_{\!\textrm{RF}}\bF^+_{\!\textrm{RF}}
\end{align}
where the second equality in \eqref{C2} holds due to the following equation
\begin{align}
\mathrm{d}\big(\bA^{\!-1}\big)\!=\!-\bA^{\!-1}\mathrm{d}\bA\bA^{\!-1}.
\end{align}
Inserting \eqref{C2} into \eqref{C1}, we have
\begin{align}\label{C3}
\mathrm{d}f(\bF_{\!\textrm{RF}})&\!\triangleq\!\trace\big(\mathrm{d}\bF_{\!\textrm{RF}}^H\nabla_{\!\bF_{\!\textrm{RF}}} f(\bF_{\!\textrm{RF}})\!+\!\nabla_{\!\bF_{\!\textrm{RF}}} f(\bF_{\!\textrm{RF}})^{H}\mathrm{d}\bF_{\!\textrm{RF}}\big)\\
&\!=\!-\trace\big(\mathrm{d}\bF_{\!\textrm{RF}}^H\bZ_1\bF_{\!\mathrm{opt}}\bZ_2^H\big)\!-\!\trace\big(\bZ_2\bF_{\!\mathrm{opt}}^H\bZ_1^H\mathrm{d}\bF_{\!\textrm{RF}}\big)
\end{align}
where $\bZ_1\!=\!\bI\!-\!\bF_{\!\textrm{RF}}\bF^+_{\!\textrm{RF}}$ and $\bZ_2\!=\!\bF^+_{\!\textrm{RF}}\bF_{\!\mathrm{opt}}$. Thus the complex gradient matrix of $f(\bF_{\!\textrm{RF}})$ is $\nabla_{\!\bF_{\!\textrm{RF}}} f(\bF_{\!\textrm{RF}})\!=\!-\bZ_1\bF_{\!\mathrm{opt}}\bZ_2^H$.

Next, we compute the Hessian matrix $\mathcal{CH}_{\bF_{\!\textrm{RF}}} f(\bF_{\!\textrm{RF}})$. Since $\mathcal{CH}_{\bF_{\!\textrm{RF}}} f(\bF_{\!\textrm{RF}})$ contains four blocks, we first determine $\mathcal{H}_{\bF_{\!\textrm{RF}},\bF_{\!\textrm{RF}}^*} f(\bF_{\!\textrm{RF}})$ and $\mathcal{H}_{\bF_{\!\textrm{RF}}^*,\bF_{\!\textrm{RF}}^*} f(\bF_{\!\textrm{RF}})$. According to the definition
\begin{align}\label{C4}
\mathrm{vec}\big[\mathrm{d} \nabla_{\!\bF_{\!\textrm{RF}}} f(\bF_{\!\textrm{RF}})\big]\!\triangleq & \mathcal{H}_{\bF_{\!\textrm{RF}},\bF_{\!\textrm{RF}}^*} f(\bF_{\!\textrm{RF}})\mathrm{vec}(\mathrm{d}\bF_{\!\textrm{RF}})\nonumber\\
&\!+\!\mathcal{H}_{\bF_{\!\textrm{RF}}^*,\bF_{\!\textrm{RF}}^*} f(\bF_{\!\textrm{RF}})\mathrm{vec}(\mathrm{d}\bF_{\!\textrm{RF}}^*)
\end{align}
we obtain $\mathcal{H}_{\bF_{\!\textrm{RF}},\bF_{\!\textrm{RF}}^*} f(\bF_{\!\textrm{RF}})$ and $\mathcal{H}_{\bF_{\!\textrm{RF}}^*,\bF_{\!\textrm{RF}}^*} f(\bF_{\!\textrm{RF}})$ through computing the differential of $\nabla_{\!\bF_{\!\textrm{RF}}} f(\bF_{\!\textrm{RF}})$:
\begin{align}
\mathrm{d} \nabla_{\!\bF_{\!\textrm{RF}}} f(\bF_{\!\textrm{RF}})\!=\! -\mathrm{d}\bZ_1\bF_{\!\mathrm{opt}}\bZ_2^H-\bZ_1\bF_{\!\mathrm{opt}}\mathrm{d}\bZ_2^H.
\end{align}
where
\begin{align}\label{C5}
\mathrm{d}\bZ_1\!=\!-\mathrm{d}\bF_{\!\textrm{RF}}\bF_{\!\textrm{RF}}^{+}\!-\!\bF_{\!\textrm{RF}}\mathrm{d}\bF_{\!\textrm{RF}}^{+},\;\mathrm{d}\bZ_2^H&\!=\!\bF_{\!\mathrm{opt}}^H(\mathrm{d}\bF^+_{\!\textrm{RF}})^H.
\end{align}
Inserting $\mathrm{d}\bF_{\!\textrm{RF}}^{+}$ in \eqref{C2} into \eqref{C5}, $\mathrm{d} \nabla_{\!\bF_{\!\textrm{RF}}} f(\bF_{\!\textrm{RF}})$ can be expressed as
\begin{align}
\mathrm{d} \nabla_{\!\bF_{\!\textrm{RF}}} f(\bF_{\!\textrm{RF}})\!=& \bZ_1\mathrm{d}\bF_{\!\textrm{RF}}\bZ_2\bZ_2^H\!+\!(\bF_{\!\textrm{RF}}^{+})^H\mathrm{d}\bF_{\!\textrm{RF}}^H\bZ_1\bF_{\!\mathrm{opt}}\bZ_2^H\nonumber\\ &\!-\!\bZ_1\bF_{\!\mathrm{opt}}\bF_{\!\mathrm{opt}}^H\bZ_1^H\mathrm{d}\bF_{\!\textrm{RF}}(\bF_{\!\textrm{RF}}^H\bF_{\!\textrm{RF}})^{-1} \nonumber\\
& +\bZ_1\bF_{\!\mathrm{opt}}\bZ_2^H\mathrm{d}\bF_{\!\textrm{RF}}^H(\bF^+_{\!\textrm{RF}})^H.
\end{align}
Then we vectorize $\mathrm{d} \nabla_{\!\bF_{\!\textrm{RF}}} f(\bF_{\!\textrm{RF}})$ using the formula $\mathrm{vec}(\bA\bX\bB)\!=\!\big(\bB^T\!\otimes\! \bA\big)\mathrm{vec}(\bX)$:
\begin{align}
\mathrm{vec}&\big[\bZ_1\mathrm{d}\bF_{\!\textrm{RF}}\bZ_2\bZ_2^H\big]\nonumber\\
&\!=\!\mathrm{vec}\big[(\bZ_1\bF_{\!\mathrm{opt}}\bZ_2^H)^T\!\otimes\!(\bF_{\!\textrm{RF}}^{+})^H\big]\mathrm{vec}(\mathrm{d}\bF_{\!\textrm{RF}})
\end{align}
\begin{align}
\mathrm{vec}&\big[\bZ_1\bF_{\!\mathrm{opt}}\bF_{\!\mathrm{opt}}^H\bZ_1^H\mathrm{d}\bF_{\!\textrm{RF}}(\bF_{\!\textrm{RF}}^H\bF_{\!\textrm{RF}})^{-1}\big]\nonumber\\
&\!=\!\mathrm{vec}\Big[\big[(\bF_{\!\textrm{RF}}^H\bF_{\!\textrm{RF}})^{-1}\big]^T\!\otimes\!\bZ_1\bF_{\!\mathrm{opt}}\bF_{\!\mathrm{opt}}^H\bZ_1^H\Big]\mathrm{vec}(\mathrm{d}\bF_{\!\textrm{RF}})
\end{align}
\begin{align}
\mathrm{vec}&\big[(\bF_{\!\textrm{RF}}^{+})^H\mathrm{d}\bF_{\!\textrm{RF}}^H\bZ_1\bF_{\!\mathrm{opt}}\bZ_2^H\big]\nonumber\\
&\!=\!\Big[(\bZ_1\bF_{\!\mathrm{opt}}\bZ_2^H)^T\!\otimes\!(\bF_{\!\textrm{RF}}^{+})^H\Big]\bK_{N_{\mathrm{t}},N_\mathrm{rf}}\mathrm{vec}(\mathrm{d}\bF_{\!\textrm{RF}}^*)
\end{align}
\begin{align}
\mathrm{vec}&\big[\bZ_1\bF_{\!\mathrm{opt}}\bZ_2^H\mathrm{d}\bF_{\!\textrm{RF}}^H(\bF^+_{\!\textrm{RF}})^H\big]\nonumber\\
&\!=\!\Big[(\bF_{\!\textrm{RF}}^{+})^*\!\otimes\!\bZ_1\bF_{\!\mathrm{opt}}\bZ_2^H\Big]\bK_{N_{\mathrm{t}},N_\mathrm{rf}}\mathrm{vec}(\mathrm{d}\bF_{\!\textrm{RF}}^*)\label{C12}
\end{align}
where $\bK_{N_{\mathrm{t}},N_\mathrm{rf}}$ is the commutation matrix such that $\mathrm{vec}(\mathrm{d}\bF_{\!\textrm{RF}}^H)\!=\! \bK_{N_{\mathrm{t}},N_\mathrm{rf}}\mathrm{vec}(\mathrm{d}\bF_{\!\textrm{RF}}^*)$. Then we can obtain
\begin{align}
\mathcal{H}_{\bF_{\!\textrm{RF}},\bF_{\!\textrm{RF}}^*} f(\bF_{\!\textrm{RF}})\!=&(\bZ_2\bZ_2^H)^T\!\otimes\! \bZ_1\nonumber\\
&\!-\!\big[(\bF_{\!\textrm{RF}}^H\bF_{\!\textrm{RF}})^{-1}\big]^T\!\otimes\!\bZ_1\bF_{\!\mathrm{opt}}\bF_{\!\mathrm{opt}}^H\bZ_1^H
\end{align}
\begin{align}
\mathcal{H}_{\bF_{\!\textrm{RF}}^*,\bF_{\!\textrm{RF}}^*} f(\bF_{\!\textrm{RF}})\!=&(\bZ_1\bF_{\!\mathrm{opt}}\bZ_2^H)^T\!\otimes\!(\bF_{\!\textrm{RF}}^{+})^H\bK_{N_{\mathrm{t}},N_\mathrm{rf}}\nonumber\\
&\!+\!(\bF_{\!\textrm{RF}}^{+})^*\!\otimes\!\bZ_1\bF_{\!\mathrm{opt}}\bZ_2^H\bK_{N_{\mathrm{t}},N_\mathrm{rf}}.
\end{align}

The remaining two blocks $\mathcal{H}_{\bF_{\!\textrm{RF}}^*,\bF_{\!\textrm{RF}}} f(\bF_{\!\textrm{RF}})$ and $\mathcal{H}_{\bF_{\!\textrm{RF}},\bF_{\!\textrm{RF}}} f(\bF_{\!\textrm{RF}})$ can be obtained via $\mathcal{H}_{\bF_{\!\textrm{RF}},\bF_{\!\textrm{RF}}^*} f(\bF_{\!\textrm{RF}})$ and $\mathcal{H}_{\bF_{\!\textrm{RF}}^*,\bF_{\!\textrm{RF}}^*} f(\bF_{\!\textrm{RF}})$. Since $\frac{\partial f(\bF_{\!\textrm{RF}})}{\partial \bF_{\!\textrm{RF}}}\!=\![\nabla_{\!\bF_{\!\textrm{RF}}} f(\bF_{\!\textrm{RF}})]^*$, $\mathrm{vec}\Big[\mathrm{d} \frac{\partial f(\bF_{\!\textrm{RF}})}{\partial \bF_{\!\textrm{RF}}}\Big]$ can be expressed as
\begin{align}
\mathrm{vec}\Bigg[\mathrm{d} \frac{\partial f(\bF_{\!\textrm{RF}})}{\partial \bF_{\!\textrm{RF}}}\Bigg]&\!\triangleq\!\mathcal{H}_{\bF_{\!\textrm{RF}}^*,\bF_{\!\textrm{RF}}} f(\bF_{\!\textrm{RF}})\mathrm{vec}(\mathrm{d}\bF_{\!\textrm{RF}}^*)\nonumber\\
&\!+\!\mathcal{H}_{\bF_{\!\textrm{RF}},\bF_{\!\textrm{RF}}}f(\bF_{\!\textrm{RF}})\mathrm{vec}(\mathrm{d}\bF_{\!\textrm{RF}})\\
&=\!\big[\mathcal{H}_{\bF_{\!\textrm{RF}},\bF_{\!\textrm{RF}}^*} f(\bF_{\!\textrm{RF}})\big]^*\mathrm{vec}(\mathrm{d}\bF_{\!\textrm{RF}}^*)\nonumber\\
&\!+\!\big[\mathcal{H}_{\bF_{\!\textrm{RF}}^*,\bF_{\!\textrm{RF}}^*} f(\bF_{\!\textrm{RF}})\big]^*\mathrm{vec}(\mathrm{d}\bF_{\!\textrm{RF}}).
\end{align}
As a consequence, one can obtain
\begin{align}
&\mathcal{H}_{\bF_{\!\textrm{RF}}^*,\bF_{\!\textrm{RF}}} f(\bF_{\!\textrm{RF}})\!=\!\big[\mathcal{H}_{\bF_{\!\textrm{RF}},\bF_{\!\textrm{RF}}^*} f(\bF_{\!\textrm{RF}})\big]^*\\
&\mathcal{H}_{\bF_{\!\textrm{RF}},\bF_{\!\textrm{RF}}} f(\bF_{\!\textrm{RF}})\!=\!\big[\mathcal{H}_{\bF_{\!\textrm{RF}}^*,\bF_{\!\textrm{RF}}^*} f(\bF_{\!\textrm{RF}})\big]^*.
\end{align}
This completes the proof.
\end{IEEEproof}

\begin{IEEEproof}[Proof of Theorem 3]
We first rewrite $\psi(\bm{\Phi}_\textrm{RF})$ as the composition of $f(\bF_{\!\textrm{RF}})$ and $\bF_{\!\textrm{RF}}(\bm{\Phi}_\textrm{RF})$, i.e.,
\begin{align}
\psi(\bm{\Phi}_\textrm{RF})\!=\!f\big[\bF_{\!\textrm{RF}}(\bm{\Phi}_\textrm{RF})\big].
\end{align}
Using the chain rule in differentiation, the differential of $\psi(\mathbf{\bar{\Phi}})$ is
\begin{align}\label{D1}
\mathrm{d}[\psi(\bm{\Phi}_\textrm{RF})]\!=\!\trace\big[\nabla_{\!\bF_{\!\textrm{RF}}} & f(\bF_{\!\textrm{RF}})^{H}\mathrm{d}\bF_{\!\textrm{RF}}(\bm{\Phi}_\textrm{RF})\nonumber\\
&\!+\!\mathrm{d}\bF_{\!\textrm{RF}}(\bm{\Phi}_\textrm{RF})^{H}\nabla_{\!\bF_{\!\textrm{RF}}} f(\bF_{\!\textrm{RF}})\big].
\end{align}
Inserting $\mathrm{d}\bF_{\!\textrm{RF}}(\bm{\Phi}_\textrm{RF})\!=\!j\bF_{\!\textrm{RF}}\!\circ\!\mathrm{d}\bm{\Phi}_\textrm{RF}$ into \eqref{D1}, $\mathrm{d}[\psi(\bm{\Phi}_\textrm{RF})]$ is expressed as
\begin{align}
\mathrm{d}[\psi(\bm{\Phi}_\textrm{RF})]&\!=\!j\trace\big[\nabla_{\!\bF_{\!\textrm{RF}}} f(\bF_{\!\textrm{RF}})^{H}(\bF_{\!\textrm{RF}}\!\circ\!\mathrm{d}\bm{\Phi}_\textrm{RF})\nonumber\\
&\!-\!(\bF_{\!\textrm{RF}}\!\circ\!\mathrm{d}\bm{\Phi}_\textrm{RF})^{H}\nabla_{\!\bF_{\!\textrm{RF}}} f(\bF_{\!\textrm{RF}})\big]\\
&\!=\!j\trace\big[(\nabla_{\!\bF_{\!\textrm{RF}}} f(\bF_{\!\textrm{RF}})^{*}\!\circ\!\bF_{\!\textrm{RF}})^{T}\mathrm{d}\bm{\Phi}_\textrm{RF}\nonumber\\
&\!-\!(\nabla_{\!\bF_{\!\textrm{RF}}} f(\bF_{\!\textrm{RF}})\!\circ\!\bF_{\!\textrm{RF}}^{*})^{T}\mathrm{d}\bm{\Phi}_\textrm{RF}\big] \label{D2}
\end{align}
where \eqref{D2} holds due to the following equality
\begin{align}
\trace\big[\bA^{\!T}(\bB\!\circ\!\bC)\big]\!=\!\trace\big[(\bA\!\circ\!\bB)^{T}\bC\big].
\end{align}
Then the gradient of $\psi(\bm{\Phi}_\textrm{RF})$ can be obtained from \eqref{D2}:
\begin{align}
\nabla \psi(\bm{\Phi}_\textrm{RF})&\!=\!j\nabla_{\!\bF_{\!\textrm{RF}}} f(\bF_{\!\textrm{RF}})^{*}\!\circ\!\bF_{\!\textrm{RF}}\!-\!j\nabla_{\!\bF_{\!\textrm{RF}}} f(\bF_{\!\textrm{RF}})\!\circ\!\bF_{\!\textrm{RF}}^{*}\\
&\!=\!2\Im\big[\nabla_{\!\bF_{\!\textrm{RF}}} f(\bF_{\!\textrm{RF}})\!\circ\! \bF_{\!\textrm{RF}}^*\big].
\end{align}

Next, we compute the Hessian of $\psi(\bm{\Phi}_\textrm{RF})$. According to the definition
\begin{align}
\mathrm{vec}\big[\mathrm{d}\nabla \psi(\bm{\Phi}_\textrm{RF})\big]\!\triangleq\!\nabla^2 \psi(\bm{\Phi}_\textrm{RF})\mathrm{vec}(\mathrm{d}\bm{\Phi}_\textrm{RF})
\end{align}
we can obtain $\nabla^2 \psi(\bm{\Phi}_\textrm{RF})$ by computing the differential of $\mathrm{vec}[\nabla \psi(\bm{\Phi}_\textrm{RF})]$:
\begin{align}
\mathrm{vec}\big[\mathrm{d}\nabla \psi(\bm{\Phi}_\textrm{RF})\big]\!=\!2\Im\big\{\mathrm{d}\big(\mathrm{vec}\big[\nabla_{\!\bF_{\!\textrm{RF}}} f(\bF_{\!\textrm{RF}})\!\circ\! \bF_{\!\textrm{RF}}^*\big]\big)\big\}.
\end{align}
Using the product rule in differentiation, $\mathrm{d}\big(\mathrm{vec}\big[\nabla_{\!\bF_{\!\textrm{RF}}} f(\bF_{\!\textrm{RF}})\!\circ\! \bF_{\!\textrm{RF}}^*\big]\big)$ is given by
\begin{align}\label{D3}
\mathrm{d}\big(\mathrm{vec}\big[\nabla_{\!\bF_{\!\textrm{RF}}} f(\bF_{\!\textrm{RF}})\!\circ\! \bF_{\!\textrm{RF}}^*\big]\big)\!=&\mathrm{vec}\big[\mathrm{d}\nabla_{\!\bF_{\!\textrm{RF}}} f(\bF_{\!\textrm{RF}})\big]\!\circ\!\mathrm{vec}(\bF_{\!\textrm{RF}}^*)\nonumber\\
&\!+\!\mathrm{vec}\big[\nabla_{\!\bF_{\!\textrm{RF}}} f(\bF_{\!\textrm{RF}})\big]\!\circ\!\mathrm{vec}(\mathrm{d}\bF_{\!\textrm{RF}}^*)
\end{align}
where
\begin{align}\label{D4}
\mathrm{vec}\big[\mathrm{d}\nabla_{\!\bF_{\!\textrm{RF}}} f(\bF_{\!\textrm{RF}})\big]\!=&\mathcal{H}_{\bF_{\!\textrm{RF}},\bF_{\!\textrm{RF}}^*} f(\bF_{\!\textrm{RF}})\mathrm{vec}(\mathrm{d}\bF_{\!\textrm{RF}})\nonumber\\
&\!+\!\mathcal{H}_{\bF_{\!\textrm{RF}}^*,\bF_{\!\textrm{RF}}^*} f(\bF_{\!\textrm{RF}})\mathrm{vec}(\mathrm{d}\bF_{\!\textrm{RF}}^*)
\end{align}
\vspace{-0.7cm}
\begin{align}
&\mathrm{vec}(\mathrm{d}\bF_{\!\textrm{RF}})\!=\!j\mathrm{vec}(\bF_{\!\textrm{RF}})\!\circ\!\mathrm{vec}(\mathrm{d}\bm{\Phi}_\textrm{RF})\\
&\mathrm{vec}(\mathrm{d}\bF_{\!\textrm{RF}}^*)\!=\!-j\mathrm{vec}(\bF_{\!\textrm{RF}}^*)\!\circ\!\mathrm{vec}(\mathrm{d}\bm{\Phi}_\textrm{RF})
\end{align}

Inserting the equations in \eqref{D4} into \eqref{D3}, $\mathrm{vec}\big[\mathrm{d}\nabla \psi(\bm{\Phi}_\textrm{RF})\big]$ can be rewritten as
\begin{align}
\mathrm{vec}&\big[\mathrm{d}\nabla \psi(\bm{\Phi}_\textrm{RF})\big]\nonumber\\
&\!=\!\big\{2\Re(\bM)\!-\!2\diag\big(\mathrm{vec}\big[\Re(\bG)\big]\big)\big\}\mathrm{vec}(\mathrm{d}\bm{\Phi}_\textrm{RF})
\end{align}
where $\bG\!=\!\nabla_{\!\bF_{\!\textrm{RF}}} f(\bF_{\!\textrm{RF}})\!\circ\! \bF_{\!\textrm{RF}}^*$, and $\bM\!=\![\mathcal{H}_{\bF_{\!\textrm{RF}},\bF_{\!\textrm{RF}}^*} f(\bF_{\!\textrm{RF}})]\!\circ\!\mathrm{vec}(\bF_{\!\textrm{RF}}^*)\mathrm{vec}(\bF_{\!\textrm{RF}})^T\!-\![\mathcal{H}_{\bF_{\!\textrm{RF}}^*,\bF_{\!\textrm{RF}}^*} f(\bF_{\!\textrm{RF}})]\circ\!\mathrm{vec}(\bF_{\!\textrm{RF}}^*)\mathrm{vec}(\bF_{\!\textrm{RF}})^H$. Therefore, the Hessian matrix $\nabla^2 \psi(\bm{\Phi}_\textrm{RF})$ is given by
\begin{align}
\nabla^2 \psi(\bm{\Phi}_\textrm{RF})\!=\!2\Re(\bM)\!-\!2\diag\big(\mathrm{vec}\big[\Re(\bG)\big]\big).
\end{align}
This completes the proof.
\end{IEEEproof}
\bibliographystyle{IEEEtran}
\bibliography{IEEEabrv,reference}

\end{document}

%% file: mathdef.tex
\def\BD{\begin{displaymath}}
\def\BE{\begin{equation}}
\def\BEA{\begin{eqnarray}}
\def\BEAs{\begin{eqnarray*}}
\def\ED{\end{displaymath}}
\def\EE{\end{equation}}
\def\EEA{\end{eqnarray}}
\def\EEAs{\end{eqnarray*}}

\def\bA{{\bf A}}
\def\ba{{\bf a}}
\def\bB{{\bf B}}

\def\bC{{\bf C}}

\def\bF{{\bf F}}
\def\bG{{\bf G}}

\def\bH{{\bf H}}

\def\bI{{\bf I}}

\def\bK{{\bf K}}

\def\bM{{\bf M}}

\def\bn{{\bf n}}

\def\bQ{{\bf Q}}

\def\bR{{\bf R}}
\def\br{{\bf r}}
\def\bS{{\bf S}}
\def\bs{{\bf s}}

\def\bU{{\bf U}}
\def\bu{{\bf u}}
\def\bV{{\bf V}}

\def\bX{{\bf X}}
\def\bx{{\bf x}}

\def\by{{\bf y}}
\def\bZ{{\bf Z}}

\def\b_eta{\mbox{\boldmath $\eta$}}

\def\bnu{\mbox{\boldmath $\nu$}}

\def\diag { {\rm diag} }

\def\trace{ {\rm tr} }

\def\L2{L^{2}[-{\pi \over 2} , {\pi \over 2}]}

\def\bnu{\bnu}

%% file: Hybrid_Precoding.bbl
\begin{thebibliography}{10}
\providecommand{\url}[1]{#1}
\csname url@samestyle\endcsname
\providecommand{\newblock}{\relax}
\providecommand{\bibinfo}[2]{#2}
\providecommand{\BIBentrySTDinterwordspacing}{\spaceskip=0pt\relax}
\providecommand{\BIBentryALTinterwordstretchfactor}{4}
\providecommand{\BIBentryALTinterwordspacing}{\spaceskip=\fontdimen2\font plus
\BIBentryALTinterwordstretchfactor\fontdimen3\font minus
  \fontdimen4\font\relax}
\providecommand{\BIBforeignlanguage}[2]{{%
\expandafter\ifx\csname l@#1\endcsname\relax
\typeout{** WARNING: IEEEtran.bst: No hyphenation pattern has been}%
\typeout{** loaded for the language `#1'. Using the pattern for}%
\typeout{** the default language instead.}%
\else
\language=\csname l@#1\endcsname
\fi
#2}}
\providecommand{\BIBdecl}{\relax}
\BIBdecl

\bibitem{el2014spatially}
O.~El~Ayach, S.~Rajagopal, S.~Abu-Surra, Z.~Pi, and R.~W. Heath, ``{Spatially
  sparse precoding in millimeter wave MIMO systems},'' \emph{{IEEE} Trans.
  Wireless Commun.}, vol.~13, no.~3, pp. 1499--1513, Mar. 2014.

\bibitem{zhang2014achieving}
E.~Zhang and C.~Huang, ``{On achieving optimal rate of digital precoder by
  RF-baseband codesign for MIMO systems},'' in \emph{Proc. 80th IEEE Veh.
  Technol. Conf. (VTC Fall)}.\hskip 1em plus 0.5em minus 0.4em\relax Vancouver,
  BC, Sept. 2014, pp. 1--5.

\bibitem{yu2016alternating}
X.~Yu, J.-C. Shen, J.~Zhang, and K.~B. Letaief, ``{Alternating minimization
  algorithms for hybrid precoding in millimeter wave MIMO systems},''
  \emph{{IEEE} J. Sel. Topics Signal Process.}, vol.~10, no.~3, pp. 485--500,
  Apr. 2016.

\bibitem{sohrabi2016hybrid}
F.~Sohrabi and W.~Yu, ``{Hybrid digital and analog beamforming design for
  large-scale antenna arrays},'' \emph{{IEEE} J. Sel. Topics Signal Process.},
  vol.~10, no.~3, pp. 501--513, Apr. 2016.

\bibitem{rusu2016low}
C.~Rusu, R.~M{\`e}ndez-Rial, N.~Gonz{\'a}lez-Prelcic, and R.~W. Heath, ``{Low
  complexity hybrid precoding strategies for millimeter wave communication
  systems},'' \emph{{IEEE} Trans. Wireless Commun.}, vol.~15, no.~12, pp.
  8380--8393, Dec. 2016.

\bibitem{rajashekar2016hybrid}
R.~Rajashekar and L.~Hanzo, ``{Hybrid beamforming in mm-wave MIMO systems
  having a finite input alphabet},'' \emph{{IEEE} Trans. Commun.}, vol.~64,
  no.~8, pp. 3337--3349, Aug. 2016.

\bibitem{rajashekar2017iterative}
``Iterative matrix decomposition aided block diagonalization for mm-wave
  multiuser mimo systems.''

\bibitem{guo2005mutual}
D.~Guo, S.~Shamai, and S.~Verd{\'u}, ``{Mutual information and minimum
  mean-square error in Gaussian channels},'' \emph{{IEEE} Trans. Inf. Theory},
  vol.~51, no.~4, pp. 1261--1282, Apr. 2005.

\bibitem{lozano2006optimum}
A.~Lozano, A.~M. Tulino, and S.~Verd{\'u}, ``{Optimum power allocation for
  parallel Gaussian channels with arbitrary input distributions},''
  \emph{{IEEE} Trans. Inf. Theory}, vol.~52, no.~7, pp. 3033--3051, Jul. 2006.

\bibitem{xiao2008mutual}
C.~Xiao and Y.~R. Zheng, ``{On the mutual information and power allocation for
  vector Gaussian channels with finite discrete inputs},'' in \emph{Proc. IEEE
  Globecom}, 2008, pp. 1--5.

\bibitem{perez2010mimo}
F.~P{\'e}rez-Cruz, M.~R. Rodrigues, and S.~Verd{\'u}, ``{MIMO Gaussian channels
  with arbitrary inputs: Optimal precoding and power allocation},''
  \emph{{IEEE} Trans. Inf. Theory}, vol.~56, no.~3, pp. 1070--1084, Mar. 2010.

\bibitem{xiao2011globally}
C.~Xiao, Y.~R. Zheng, and Z.~Ding, ``{Globally optimal linear precoders for
  finite alphabet signals over complex vector Gaussian channels},''
  \emph{{IEEE} Trans. Signal Process.}, vol.~59, no.~7, pp. 3301--3314, Jul.
  2011.

\bibitem{zeng2012low}
W.~Zeng, C.~Xiao, and J.~Lu, ``{A low-complexity design of linear precoding for
  MIMO channels with finite-alphabet inputs},'' \emph{{IEEE Wireless Commun.
  Lett.}}, vol.~1, no.~1, pp. 38--41, Feb. 2012.

\bibitem{jin2017linear}
J.~Jin, C.~Xiao, M.~Tao, and W.~Chen, ``{Linear Precoding for Fading Cognitive
  Multiple-Access Wiretap Channel With Finite-Alphabet Inputs},'' \emph{{IEEE}
  Trans. Veh. Technol.}, vol.~66, no.~4, pp. 3059--3070, Apr. 2017.

\bibitem{jin2017generalized}
J.~Jin, Y.~R. Zheng, W.~Chen, and C.~Xiao, ``{Generalized quadratic matrix
  programming: a unified framework for linear precoding with arbitrary input
  distributions},'' \emph{{IEEE} Trans. Signal Process.}, vol.~65, no.~18, pp.
  4887--4901, Sept. 2017.

\bibitem{tse2005fundamentals}
D.~Tse and P.~Viswanath, \emph{Fundamentals of wireless communication}.\hskip
  1em plus 0.5em minus 0.4em\relax Cambridge, U.K.: Cambridge Univ. Press,
  2005.

\bibitem{hjorungnes2007complex}
A.~Hjorungnes and D.~Gesbert, ``{Complex-valued matrix differentiation:
  Techniques and key results},'' \emph{{IEEE} Trans. Signal Process.}, vol.~55,
  no.~6, pp. 2740--2746, Jun. 2007.

\bibitem{raghavan2016beamforming}
V.~Raghavan, J.~Cezanne, S.~Subramanian, A.~Sampath, and O.~Koymen,
  ``{Beamforming tradeoffs for initial UE discovery in millimeter-wave MIMO
  systems},'' \emph{{IEEE} J. Sel. Topics Signal Process.}, vol.~10, no.~3, pp.
  543--559, Apr. 2016.

\bibitem{chen2015solving}
Y.~Chen and E.~Candes, ``Solving random quadratic systems of equations is
  nearly as easy as solving linear systems,'' in \emph{Proc. Adv. Neural Inf.
  Process. Syst.}, 2015.

\bibitem{li2001global}
D.-H. Li and M.~Fukushima, ``{On the global convergence of the BFGS method for
  nonconvex unconstrained optimization problems},'' \emph{{SIAM} J. Optim.},
  vol.~11, no.~4, pp. 1054--1064, May 2001.

\bibitem{boyd2004convex}
S.~Boyd and L.~Vandenberghe, \emph{{Convex Optimization}}.\hskip 1em plus 0.5em
  minus 0.4em\relax Cambridge, U.K.: Cambridge Univ. Press, 2004.

\bibitem{dauphin2014identifying}
Y.~N. Dauphin, R.~Pascanu, C.~Gulcehre, K.~Cho, S.~Ganguli, and Y.~Bengio,
  ``Identifying and attacking the saddle point problem in high-dimensional
  non-convex optimization,'' in \emph{Proc. Adv. Neural Inf. Process. Syst.},
  2014.

\bibitem{reddi2017generic}
S.~J. Reddi, M.~Zaheer, S.~Sra, B.~Poczos, F.~Bach, R.~Salakhutdinov, and A.~J.
  Smola, ``{A generic approach for escaping saddle points},''
  \emph{arXiv:1709.01434}, Sept. 2017.

\bibitem{lasserre1995trace}
J.~B. Lasserre, ``{A trace inequality for matrix product},'' \emph{{IEEE}
  Trans. Autom. Control}, vol.~40, no.~8, pp. 1500--1501, Aug. 1995.

\end{thebibliography}
